\documentclass[fleqn,usenatbib]{mnras}

\usepackage{newtxtext,newtxmath}
\usepackage{color, colortbl}
\usepackage{lineno}
\definecolor{Gray}{gray}{0.9}

\usepackage[T1]{fontenc}

\DeclareRobustCommand{\VAN}[3]{#2}
\let\VANthebibliography\thebibliography
\def\thebibliography{\DeclareRobustCommand{\VAN}[3]{##3}\VANthebibliography}

\usepackage{graphicx}	
\usepackage{amsmath}	
\usepackage{ulem}

\title[Two-fluid jet at a magnetic null point]{Numerical simulations of a two-fluid jet at a magnetic null point in a solar arcade}

\author[Gonz\'alez-Avil\'es et al.]{J. J. Gonz\'alez-Avil\'es$^{1}$\thanks{E-mail: jjgonzalez@igeofisica.unam.mx}
K. Murawski$^{2}$
and T. V. Zaqarashvili$^{3,4,5}$
\\
$^{1}$Investigadores por M\'exico-CONACYT, Servicio de Clima Espacial M\'exico, Laboratorio Nacional de Clima Espacial, Instituto de Geof\'isica, Unidad Michoac\'an, \\ Universidad Nacional Aut\'onoma de M\'exico, 58190 Morelia, Michoac\'an, M\'exico\\
$^{2}$ Institute of Physics, University of Maria Curie-Sk{\l}odowska, Pl. M. Curie-Sk{\l}odowskiej 5, PL-20-031 Lublin, Poland \\
$^{3}$ IGAM, Institute f\"ur Physik, University of Graz, Universit\"atsplatz 5, A-8010 Graz, Austria \\
$^{4}$ Ilia State University, Cholokashvili ave 5/3, Tbilisi, Georgia \\
$^{5}$ Abastumani Astrophysical Observatory, Mount Kanobili, Abastumani, Georgia \\
}

\date{Accepted 2022 July 15. Received 2022 June 21; in original form 2021 November 05}

\pubyear{2022}

\begin{document}
\label{firstpage}
\pagerange{\pageref{firstpage}--\pageref{lastpage}}
\maketitle

\begin{abstract}

We study the formation and evolution of jets in the solar atmosphere using numerical simulations of partially ionized plasma. The two-fluid magnetohydrodynamic equations with ion+electron and neutral hydrogen components are used in two-dimensional (2D) Cartesian geometry. Numerical simulations show that a localized nonlinear Gaussian pulse of ion and neutral pressures initially launched from the magnetic null point of a potential arcade located below the transition region quickly develops into a shock due to the decrease of density with height. The shock propagates upwards into the solar corona and lifts the cold and dense chromospheric plasma behind in the form of a collimated jet with an inverted-Y shape. The inverted-Y shape of jets is connected with the topology of a magnetic null point. The pulse also excites a nonlinear wake in the chromosphere, which leads to quasi-periodic secondary shocks. The secondary shocks lift the chromospheric plasma upwards and create quasi-periodic jets in the lower corona. Ion and neutral fluids show generally similar behavior, but their relative velocity is higher near the upper part of jets, which leads to enhanced temperature or heating due to ion-neutral collisions. Simulations of jets with inverted-Y shape and their heating may explain the properties of some jets observed in the solar atmosphere.

\end{abstract}

\begin{keywords}
(magnetohydrodynamics) MHD -- methods: numerical -- Sun: atmosphere -- Sun: chromosphere
\end{keywords}


\onecolumn
\section{Introduction}
\label{sec:Introduction}

Nowadays, the knowledge of the solar atmosphere properties has significantly improved, but it is still challenging to understand its dynamics and associated phenomena fully. In particular, the fine and small-scale plasma structures still pose many open questions to answer \citep{Judge_2006,Lipartito_et_al_2014,Kiss_et_al_2017}. Among these fine structures are spicules, which are spiky-like gas jets mainly observed in the chromosphere \citep{Secchi_1877, Sterling_2000, De_Pontieu_et_al_2004}, and the inverted-Y shape jet events first identified by \citet{Shibata_et_al_2007}, which could manifest as bright regions on broadband Ca II H images \citep{Morita_et_al_2010,Nishizuka_et_al_2011,Singh_et_al_2012}. In the context of numerical magnetohydrodynamic (MHD) simulations of inverted-Y shape jets, \citet{Nishizuka_et_al_2008} showed the formation of giant chromospheric anemone jets with the inverted-Y shape, as observed by Hinode. \citet{Yang_et_al_2013} found that 2.5D numerical MHD simulations reproduced many observed chromospheric anemone jets with inverted-Y shapes. The formation of such jets are commonly related to signatures of magnetic reconnection in sunspots \citep{Tian_et_al_2018}, in the solar photosphere \citep{Nelson_et_al_2019}, in the upper chromosphere and in the TR \citep{Nishizuka_et_al_2011}. The magnetic reconnection may take place, for instance, near magnetic null points, which are expected to be rather common in different layers of the solar atmosphere \citep{Close_et_al_2004,Regnier_et_al_2008,Freed_et_al_2015}. Among the numerical simulations related to the null points, there are studies of cool and dense blobs \citep{Murawski_et_al_2011b} as well as MHD wave propagation and shocks \citep[see, e.g.,][]{Kuzma_et_al_2015, Sabri_et_al_2020, Pennicott&Cally_2021}. Besides, \cite{Smirnova_et_al_2016} performed 2D numerical simulations of jet-like structures by launching a pressure pulse from the null point of a potential magnetic arcade in the framework of the single-fluid MHD and found jets that mimic some properties of type I and type II spicules. 

The numerical modeling of small-scale jets, including spicules and inverted-Y shape jets, has been mainly developed by adopting the single-fluid MHD approach. However, in the lower solar atmosphere, which covers the region from the photosphere to the upper chromosphere, a significant fraction of neutrals must be taken into account to describe the atmosphere realistically \citep{Zaqarashvili_et_al_2011}. The effect of neutral atoms can be studied either in the framework of a single-fluid MHD model with Cowling conductivity or in the two-fluid MHD approach. In the two-fluid model ions+electrons and neutral atoms are treated as two separate fluids which are coupled by ion-neutral collisions \citep{Zaqarashvili_et_al_2011,Soler_et_al_2013,Leake_et_al_2013,Oliver_et_al_2016,Kuzma_et_al_2017b,2019SoPh..294..124A,Wojcik_et_al_2018}.

In this paper, we extend the model of \citet{Smirnova_et_al_2016} and focus our analysis on a small-scale and inverted-Y shape jet (not limited to spicules) using the two-fluid equations that include ions+electrons and neutrals. In particular, we simulate the formation of an impulsively generated jet by pressure pulses initially launched from a magnetic null point of a potential arcade. We suppose that a trigger of the pressure pulse could be magnetic reconnection, but we do not describe the details of the excitation. Instead, we aim to study the formation and evolution of the jet. 

We describe in detail the two-fluid equations, the model of the solar atmosphere, the magnetic field configuration, perturbations, and the numerical methods in Section \ref{sec:Model&methods}. In Section \ref{sec:results_of_numerical_simulations_analysis}, we present the results of numerical simulation and the analysis, and provide the final comments and conclusions in section \ref{sec:Conclusions}.   

\section{Model and methods}
\label{sec:Model&methods}

\subsection{The system of two-fluid equations}
\label{subsec:two-fluid_equations}

To model the formation and evolution of a jet in the solar atmosphere, we consider a plasma that consists of two components: ionized fluid (ions+electrons) and neutral fluid (hydrogen atoms) that interact via ion-neutral collisions \citep[see, e.g.][]{Draine_et_al_1983,Smith&Sakai_2008,Zaqarashvili_et_al_2011,Khomenko_et_al_2014,Leake_et_al_2014,Kuzma_et_al_2017b}. In particular, we write the two-fluid equations in conservative form as follows \citep[see, e.g.][]{Oliver_et_al_2016}:
\begin{eqnarray}
\frac{\partial\varrho_{i}}{\partial t} + \nabla\cdot(\varrho_{i}{\bf v}_{i})&=&0, \label{density_ions}\\
\frac{\partial\varrho_{n}}{\partial t} + \nabla\cdot(\varrho_{n}{\bf v}_{n})&=&0, \label{density_neutrals} \\
\frac{\partial(\varrho_{i}{\bf v}_{i})}{\partial t} + \nabla\cdot\left(\varrho_{i}{\bf v}_{i}{\bf v}_{i} + \left(p_{i} + \frac{{\bf B}^{2}}{2}\right){\bf I}-{\bf B}{\bf B}\right) &=& \alpha_{in}({\bf v}_{n}-{\bf v}_{i}) + \varrho_{i}{\bf g}, \label{mom_eq_ions} \\
\frac{\partial(\varrho_{n}{\bf v}_{n})}{\partial t} + \nabla\cdot\left(\varrho_{n}{\bf v}_{n}{\bf v}_{n}+p_{n}{\bf I}\right) &=& -\alpha_{in}({\bf v}_{n}-{\bf v}_{i}) + \varrho_{n}{\bf g}, \label{mom_eq_neutrals} \\
\frac{\partial E_{i}}{\partial t} + \nabla\cdot\left(\left(E_{i} + p_{i} + \frac{{\bf B}^{2}}{2}\right){\bf v}_{i}-{\bf B}({\bf v}_{i}\cdot{\bf B})\right) &=& \alpha_{in}{\bf v}_{i}\cdot({\bf v}_{n}-{\bf v}_{i}) + \varrho_{i}{\bf g}\cdot{\bf v}_{i} + Q_{i}, \label{Energy_eq_ions} \\
\frac{\partial E_{n}}{\partial t} + \nabla\cdot[(E_{n}+p_{n}){\bf v}_{n}]  &=& \alpha_{in}{\bf v}_{n}\cdot({\bf v}_{i}-{\bf v}_{n}) + \varrho_{n}{\bf g}\cdot{\bf v}_{n} + Q_{n}, \label{Energy_eq_neutrals} \\
\frac{\partial{\bf B}}{\partial t} +\nabla\cdot({\bf v}_{i}{\bf B} -{\bf B}{\bf v}_{i}) &=& {\bf 0}, \label{evolB} \\
\nabla\cdot{\bf B} &=& 0. \label{divergenceB} 
\end{eqnarray}
\noindent Here $\varrho_{i,n}$ are the mass densities of ions and neutrals, ${\bf v}_{i,n}$ represent the velocities, $p_{i,n}$ are the ion+electron and neutral gas pressures, $E_{i,n}$ are the total energy densities, $I$ is the unit matrix tensor, and ${\bf B}$ is the magnetic field which is normalized by $\sqrt{\mu}$ with $\mu$ being the magnetic permeability. In this paper we consider the ratio of specific heats $\gamma=5/3$, and the gravitational source term on the right hand side of equations (\ref{mom_eq_ions})-(\ref{Energy_eq_neutrals}) is given by ${\bf g}=(0,-g,0)$ with its magnitude $g=274.78$ m s$^{-2}$, which represents an average over the solar surface. In equations (\ref{Energy_eq_ions})-(\ref{Energy_eq_neutrals}), the total energy densities $E_{i}$ and $E_{n}$ are given as
\begin{eqnarray}
E_{i} &=& \frac{p_{i}}{\gamma-1} + \frac{\varrho_{i}|{\bf v}_{i}|^{2}}{2} + \frac{{|\bf B|}^{2}}{2}, \label{total_energy_density_ions} \\
E_{n} &=& \frac{p_{n}}{\gamma-1} + \frac{\varrho_{n}|{\bf v}_{n}|^{2}}{2}.  \label{total_energy_density_neutrals}
\end{eqnarray}
We neglect the effect of ionization, recombination, and electron dynamics. The latter is justified as the electron mass is small compared to the ion and neutral masses. The effect of the interaction between ions and neutrals depends on friction coefficient, $\alpha_{in}$, defined by assuming that the ion-neutral momentum transfer cross-section, $\sigma_{in}$, is independent of the relative velocity of the colliding particles. This assumption is valid because charge exchange interactions (for which $\sigma_{in}\approx$ constant) generally contribute more to the momentum exchange than elastic scattering. Considering these approximations, $\alpha_{in}$ is defined \citep{Braginskii_1965} as
\begin{equation}
 \alpha_{in} = \frac{4}{3}\frac{\sigma_{in}}{m_{i}+m_{n}}\sqrt{\frac{8k_{B}}{\pi}\left(\frac{T_{i}}{m_{i}}+\frac{T_{n}}{m_{n}}\right)}\varrho_{i}\varrho_{n}, 
\end{equation}
where $k_{B}$ is Boltzmann's constant, $m_{i,n}$ are the atomic masses, and $T_{i,n}$ are their temperatures \citep[see, e.g.][]{Braginskii_1965,Chapman&Cowling_1970}. Regarding the proton-hydrogen collision cross section, we use $\sigma_{in}=1.4\times 10^{-19}$ m$^{2}$ \citep{Vranjes&Krstic_2013}. The frictional interaction between ions and neutrals results in additional heat production exchange terms, $Q_{i,n}$, in equations (\ref{Energy_eq_ions}) and (\ref{Energy_eq_neutrals}). These terms are specified as \citep[see, e.g.,][]{Ballester_et_al_2018}
\begin{eqnarray}
Q_{i}= \alpha_{in}\left[\frac{1}{2}|{\bf v}_{i}-{\bf v}_{n}|^{2}+ \frac{3}{2}\frac{k_{B}}{m_{H}(\mu_{i}+\mu_{n})}(T_{n}-T_{i})\right], \label{Qi} \\
Q_{n}= \alpha_{in}\left[\frac{1}{2}|{\bf v}_{i}-{\bf v}_{n}|^{2}+\frac{3}{2}\frac{k_{B}}{m_{H}(\mu_{i}+\mu_{n})}(T_{i}-T_{n})\right]. \label{Qn}
\end{eqnarray}
Equations (\ref{density_ions})-(\ref{divergenceB}) are supplemented by the following ideal gas laws:
\begin{equation}
p_{i} = \frac{k_{B}}{m_{i}}\varrho_{i}T_{i}, \quad p_{n} = \frac{k_{B}}{m_{n}}\varrho_{n}T_{n}. \label{EOS}
\end{equation}
Here $m_{i}=m_{H}\mu_{i}$, $m_{n}=m_{H}\mu_{n}$, and $m_{H}$ represents the hydrogen mass, which is the main ingredient of the gas, and therefore $m_{n}\simeq m_{i}=m_{H}=m_{p}$ (with $m_{p}$ being the proton mass). The mean masses $\mu_{i}\approx0.58$ and $\mu_{n}\approx1.21$, which include the influence from heavier elements taken from the OPAL solar abundance table model \citep[see, e.g.][]{Vogler_et_al_2005}. 

The system of equations (\ref{density_ions})-(\ref{divergenceB}) is general and represent a 3D Cartesian dimensional space, however we limit our model to a two-dimensional (2D) case, where $y$ axis is directed vertically upwards, with transverse components $(\partial/\partial z=0)$ of velocities and magnetic field set identically to zero, i.e., $v_{iz}=v_{nz}=B_{z}=0$. As a result, Alfv\'en waves are removed from the consideration.


\subsection{Model of the solar atmosphere}
\label{subsec:model_atmosphere}
\begin{figure*}
\centering
\includegraphics[width=8.0cm,height=5.5cm]{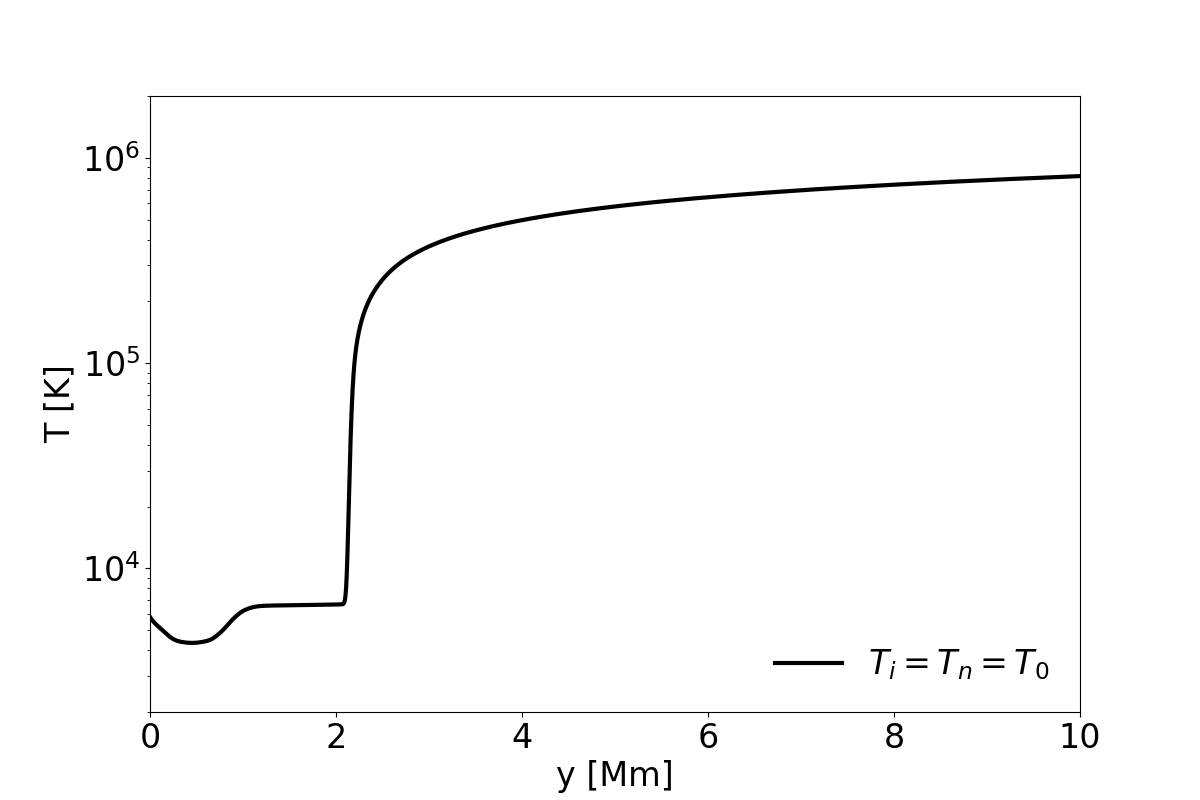}
\includegraphics[width=8.0cm,height=5.5cm]{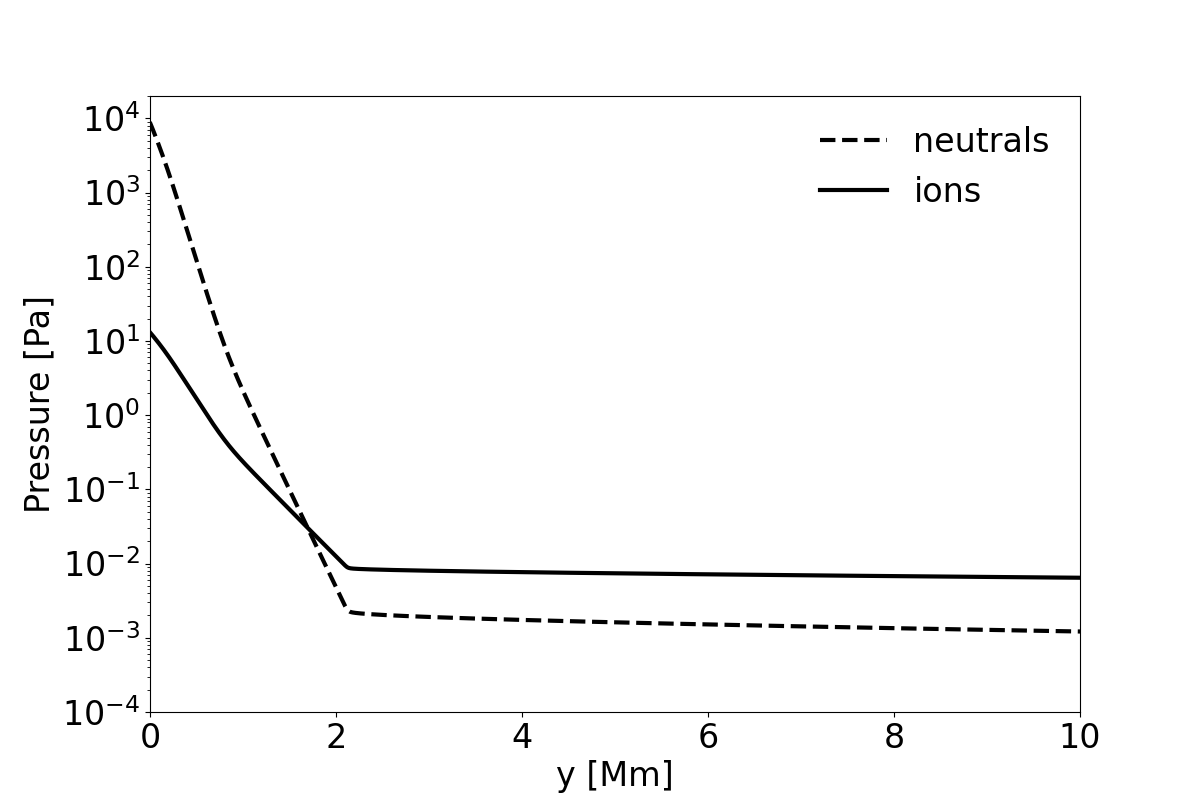}\\
\includegraphics[width=8.0cm,height=5.5cm]{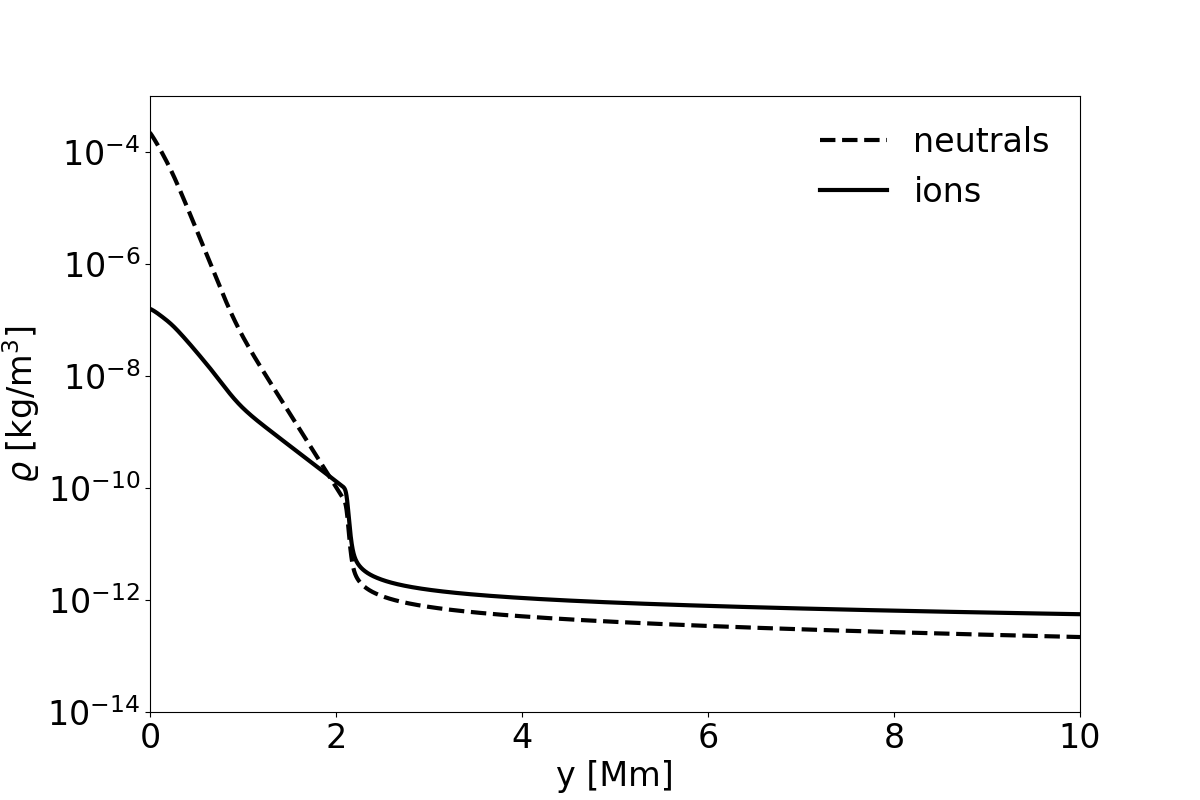}
\includegraphics[width=8.0cm,height=5.5cm]{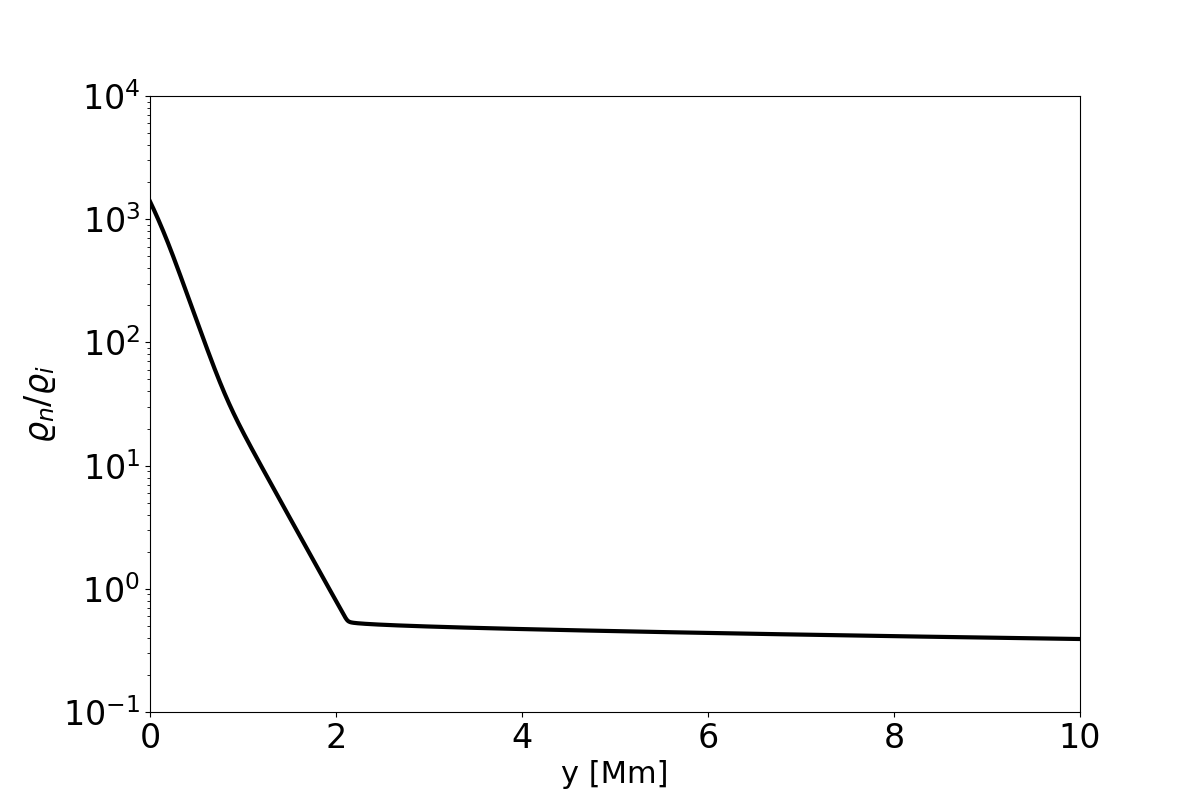}
\caption{Equilibrium profiles for temperature (top-left), gas pressures of ions and neutrals (top-right). Mass densities (bottom-left), and ionization fraction, $\varrho_{n}/\varrho_{i}$, (bottom-right) vs height $y$.}
\label{atmosphere_model}
\end{figure*}

At the initial time of simulations, we assume that the solar atmosphere is in magnetohydrostatic equilibrium, i.e., we set the ion and neutral velocities equal to zero (${\bf v_{i}=v_{n}=0}$). Additionally, the Lorentz force acts directly on the ions, and it should balance the gravity force and the gas pressure gradient, 
\begin{equation}
  (\nabla\times{\bf B})\times{\bf B}-\nabla p_{i}+\varrho_{i}{\bf g}={\bf 0}, \label{eq_hydrostatic_ions} 
\end{equation}
while for neutrals the hydrostatic equation is satisfied, 
\begin{equation}
 -\nabla p_{n} + \varrho_{n}{\bf g} = {\bf 0}. \label{eq_hydrostatic_neutrals}
\end{equation}
Considering the ideal gas laws for ions and neutrals, given by equation (\ref{EOS}), and taking into account the $y$-components of magnetohydrostatic equation (\ref{eq_hydrostatic_ions}) for a force-free magnetic field ($(\nabla\times{\bf B})\times{\bf B}={\bf 0}$) and equation (\ref{eq_hydrostatic_neutrals}), we arrive to the expressions for the equilibrium gas pressures
\begin{eqnarray}
p_{n}(y) &=& p_{0n}\exp{\left(-\int_{y_{0}}^{y}\frac{dy^{\prime}}{\Lambda_{n}(y^{\prime})}\right)}, \label{p_n} \\
p_{i}(y) &=& p_{0i}\exp{\left(-\int_{y_{0}}^{y}\frac{dy^{\prime}}{\Lambda_{i}(y^{\prime})}\right)}. \label{p_i} 
\end{eqnarray}
Here
\begin{eqnarray}
\Lambda_{i}(y) = \frac{k_{B}T_{i}(y)}{m_{H}\mu_{i}g} \quad \mbox{and} \quad \Lambda_{n}(y) = \frac{k_{B}T_{n}(y)}{m_{H}\mu_{n}g} 
\end{eqnarray}
are the pressure scale heights, and $p_{0_{i,n}}$ denote the gas pressures at the reference base-level $y_{0}=0$ Mm, i.e., at the photosphere. The pressures $p_{0_{i,n}}$ are the free parameters of the model. So, we can obtain different profiles for different values of $p_{0_{i,n}}$. In this paper, we estimate $p_{0_{i,n}}$ using the number density of ions and neutrals at $y_{0}=0$ Mm from the C7 model of \citet{AvretLoeser2008}, with the ions and neutral number densities $n_{i0}(y=y_{0})=8.397\times10^{19}$ m$^{-3}$, and $n_{n0}(y=y_{0})=1.187\times10^{23}$ m$^{-3}$. Then, we use equation (\ref{EOS}) and get $p_{0_{i}}=13.1528$ Pa and $p_{0_{n}}=8912.28$ Pa. This choice gives profiles that consistently describe the behavior of mass densities from the photosphere up to the TR that is located at $y\approx2.1$ Mm. The mass densities are given by equation (\ref{EOS}). Specifically, in this paper, we adopt the semi-empirical model of Table 26 of \citet{AvretLoeser2008} for the temperature field. We consider that temperatures of ions and neutrals are initially equal (at $t=0$ s), i.e., they are in thermal equilibrium; we simply set $T_{i}=T_{n}=T$ and show on top-left of Fig. \ref{atmosphere_model}. We display all the equilibrium profiles including the mass densities and gas pressures for ions and neutrals and the ionization fraction, $\varrho_{n}/\varrho_{i}$, for a region up to $y=10$ Mm in Fig. \ref{atmosphere_model}. Note that at the bottom of the photosphere ($y=0$ Mm), the mass density of neutrals ($\sim 10^{-4}$ kg m$^{-3}$) is higher than the mass density of ions ($\approx 10^{-7}$ kg m$^{-3}$), while at higher altitudes ($y>2.1$ Mm) the mass density of ions ($\sim10^{-12}$ kg m$^{-3}$) is higher than mas density of neutrals ($\sim10^{-14}$ kg m$^{-3}$). Moreover, gas pressure is higher for neutrals ($\sim10^{4}$ Pa) than for ions ($\sim10$ Pa) at $y=0$ Mm. However, the ion pressure ($p_i \sim 10^{-2}$ Pa) is higher than the neutral pressure ($p_n \sim 10^{-3}$ Pa) at higher altitudes. A similar relation holds for mass densities, i.e., $\varrho_i>\varrho_n$.

As a result of the adopted profiles for the mass densities, the ionization fraction ($\varrho_{n}/\varrho_{i}$) is about $10^{3}$ at $y=0$ Mm. In contrast, the ionization fraction tends to $10^{-1}$ at coronal heights ($y>2.1$ Mm), so the gas is essentially fully ionized there. Although this ionization fraction seems to be a bit too high, it could be possible to obtain a smaller ionization fraction, {\it, e.g.} of about $10^{-6}$, in the solar corona by starting with the profiles of mass densities of ions and neutrals in the C7 model. Therefore, we would obtain different profiles of temperatures for ions and neutrals. However, this would also imply that $T_{n}$ is only $100$ K at the TR \citep[see, e.g.,][]{Kuzma_et_al_2017b}. Alternatively, we could also modify the amplitude of reference pressures $p_{0_{i,n}}$ to obtain a lower ionization fraction in the solar corona, but it could not be easy to maintain the realistic ionization fraction ($\sim 10^{3}$) in the photosphere. 

In summary, by solving the hydrostatic equilibrium, we can be closer to the strongly ionized limit in the corona, but it would depart us a bit too far from the partially ionized limit at the photosphere, or vice versa, but not both simultaneously. Therefore, we decided to keep initially the same temperatures for ions and neutrals and take the pressure values at the reference base-level $y=0$ Mm from the C7 model \citep{AvretLoeser2008}. Despite the ionization fraction $\sim10^{-1}$ in the low corona, the gas is not as strongly ionized as in Fig. 1 of \citet{10.1093/mnras/staa3835}. These conditions are still reasonable as it is shown in the two-fluid simulations of jets and waves in the solar atmosphere \citep[see, e.g.,][]{Kuzma_et_al_2021,Murawski_et_al_2020,Wojcik_et_al_2019, Srivastava_et_al_2018}.
%
%
\subsection{Magnetic field configuration}
\label{subsec:model_mag_field}
\begin{figure*}
\centering
\includegraphics[width=5.0cm,height=7.0cm]{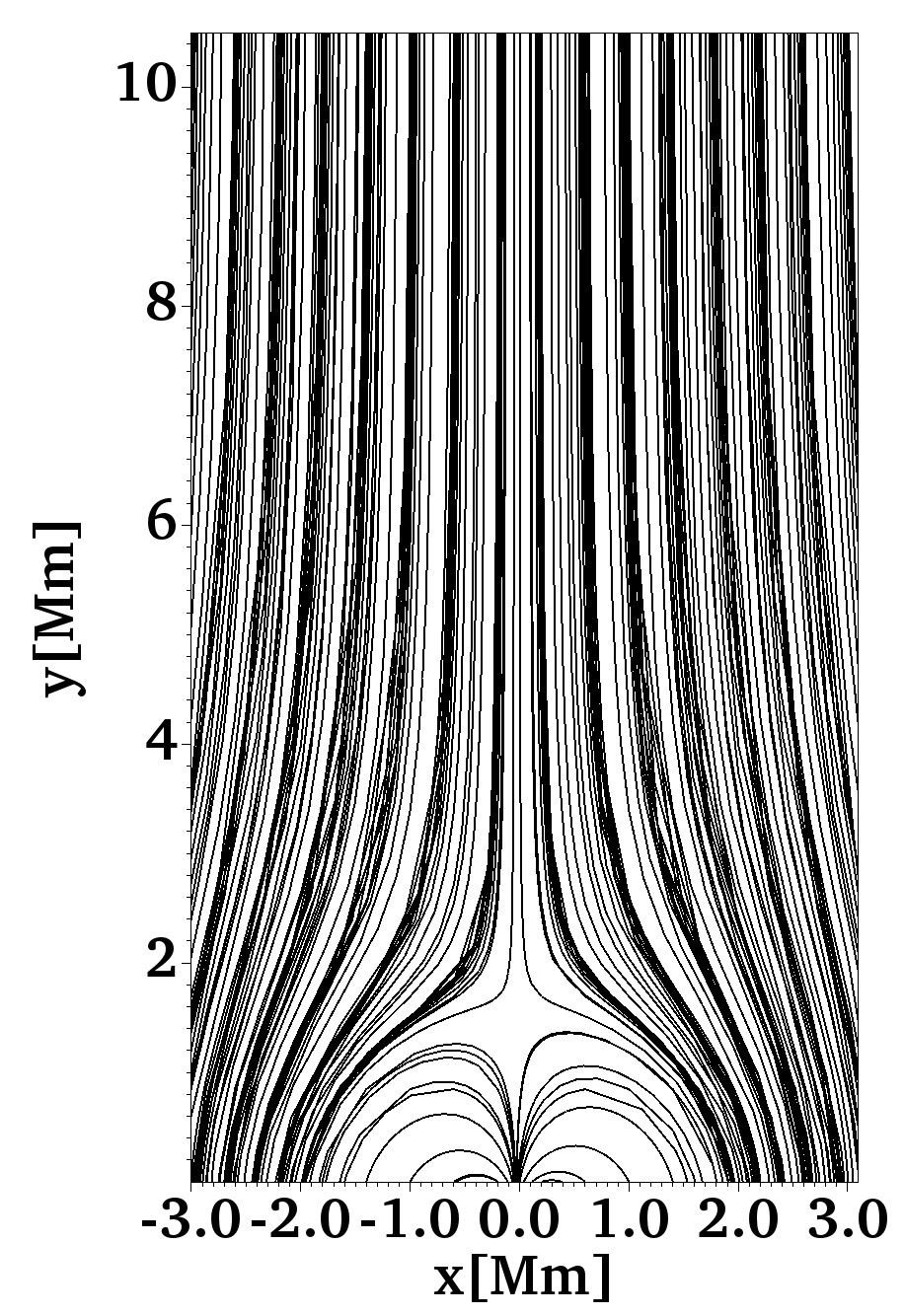}
\includegraphics[width=6.75cm,height=7.0cm]{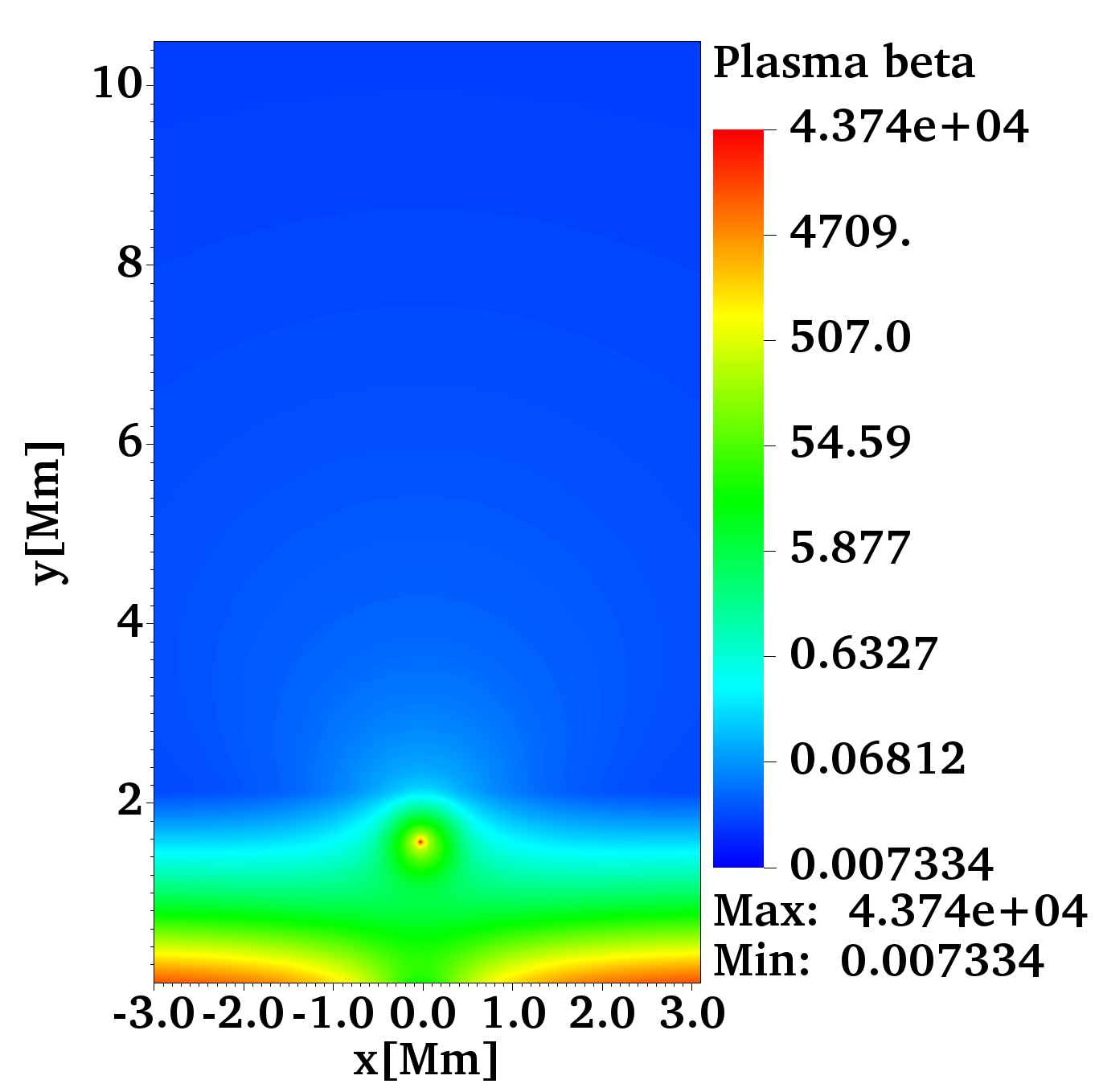}
\caption{Magnetic field lines for $b=20$ G, $h=0.3$ Mm, and $k=3.33$ Mm$^{-1}$ (left) and bulk plasma $\beta$ (right) at the initial state ($t=0$ s). In these plots the location of the null point is at $y=$1.6 Mm.}
\label{mag_field}
\end{figure*}

In a scenario of the solar atmosphere being in magnetohydrostatic equilibrium at initial time of the simulations, the magnetic field is force-free ($(\nabla\times{\bf B})\times{\bf B}={\bf 0}$) and it satisfies the divergence-free condition ($\nabla\cdot{\bf B}=0$). We additionally assume that the magnetic field is current-free ($\nabla\times{\bf B}={\bf 0}$). We define the magnetic field components in terms of magnetic flux function, $A(x,y)$, given as, 
${\bf B}=\nabla\times(A{\bf \hat{z}})$, where ${\bf\hat{z}}$ is a unit vector along the $z-$direction. The term magnetic flux function is commonly used for a 2D magnetic field to link with the definition of the reconnection rate and, particularly, to the different fluxes in the partition corresponding to the regions of the plane bounded by separatrices, which are the global stable and unstable manifolds of hyperbolic nulls points \citep[see, e.g.,][]{Yeates&Horning_2011}. Therefore, the horizontal, $B_{x}$, and vertical, $B_{y}$, magnetic field components are: 
\begin{eqnarray}
B_{x} = -\frac{\partial A(x,y)}{\partial y}, \quad 
B_{y}=   \frac{\partial A(x,y)}{\partial x}.  \label{mag_field_components}
\end{eqnarray}
Here, $A(x,y)$ is specified as in \citet{Smirnova_et_al_2016},
\begin{equation}
A(x,y) = B_{0}\frac{x}{(kx)^{2}+(k(y+h))^{2}}-bx, \label{mag_flux_function_full}
\end{equation}
with $B_{0}$ representing the strength of the magnetic field of the arcade, $h$ is the depth of the point of the magnetic singularity, $k=1/h$ is the inverse of the depth, and $b$ is the strength of a vertical ambient magnetic field. Thus, $A(x,y)$ describes the geometry of a single magnetic arcade with the null point located on the symmetry axis ($x=0$). See Fig. \ref{mag_field} (left). 

We use equation (\ref{mag_field_components}) to express the magnetic field components in the following forms:
\begin{eqnarray}
B_{x}(x,y) &=& \frac{2B_{0}x(y+h)}{k^{2}(x^{2}+(y+h)^{2})^{2}}, \label{Bx_field} \\
B_{y}(x,y) &=& \frac{B_{0}(h-x+y)(h+x+y)}{k^{2}(x^{2}+(y+h)^{2})^{2}}-b. \label{By_field}
\end{eqnarray}
According to \citet{Smirnova_et_al_2016}, for $k=1/h$, $B_{0}=B_{y}(0,0)$ has the physical meaning of the magnetic field strength at the point $(x=0,y=0)$. Along $y-$axes (for $x=0$ Mm) we have 
\begin{equation}
B_{y}(x=0,y) = \frac{B_{0}}{\left(\frac{y}{h}+1\right)^{2}}-b. 
\end{equation}
\noindent Hence, the coordinate of the null point $Y$, where $B_{y}(0,Y)=0$, can be calculated from
\begin{equation}
B_{0} = b\left(\frac{Y}{h}+1\right)^{2}. 
\label{B0_straight}
\end{equation}
In this paper, we let vary the parameters $h$, $Y$, $b$, and $B_{0}$ and consider two different values of $y$. The chosen parameters are essentially the same as in \citet{Smirnova_et_al_2016}. 

The plasma $\beta$ parameter helps estimate the ratio between ion and neutral pressures to magnetic pressure. We define the bulk plasma $\beta$ as follows:
\begin{equation}
\beta(x,y)= \frac{p_{i}(y)+p_{n}(y)}{B^{2}(x,y)/2}.   
\end{equation}
\noindent Here, the pressures $p_{i,n}(y)$ are given by equations (\ref{p_n})-(\ref{p_i}), and $B^{2}=(B_{x}^{2}+B_{y}^{2})$. The spatial profile of plasma $\beta$ is displayed on the right panel of Fig. \ref{mag_field}, where we observe that $\beta>1$ in the lower atmosphere (the photosphere and the chromosphere) and at the null point $\beta\rightarrow \infty$; Otherwise, $\beta<1$ in the solar corona ($y>2.1$ Mm). Such behavior of plasma $\beta$ is consistent to our expectations.
%
%
\subsection{Perturbations}
\label{subsec:perturbations}
We perturb the magnetohydrostatic equilibrium atmosphere, 
described in Sections \ref{subsec:model_atmosphere} and \ref{subsec:model_mag_field}, initially (at $t=0$\, s), by localized Gaussian pulses in ion and neutral gas pressures \citep[see, e.g.][]{Kuzma_et_al_2017b}, given as
\begin{equation}
p_{i,n}(x,y,t=0) = p_{i,n}(y)\left[1+ A_{p}\exp\left(-\frac{(x-x_{0})^{2}+(y-y_{0})^{2}}{w^{2}}\right)\right]. \label{perturbation}
\end{equation}
\noindent Here $A_{p}$ denotes the relative amplitude of the pulses, $(x_{0},y_{0})$ is their initial position, and $w$ their width. We set and hold fixed the value of $x_{0}$ to 0.0 Mm and let $y_{0}$ vary according to the position of the null point $Y$. Therefore, we launch the perturbation from the upper chromosphere ($y_{0}=1.6$ Mm) and above the TR, i.e., from the low corona ($y_{0}=2.2$ Mm). In our simulations, we set $w=0.3$ Mm and $A_{p}=8.5$, which are the same values as used by \citet{Smirnova_et_al_2016}. The signals launched from the null point location may mimic a post-reconnection event. The perturbations produce energy deviations to the background atmosphere at the initial time, that we estimate using the equations (\ref{Energy_eq_ions}) and (\ref{Energy_eq_neutrals}). Therefore, the deviations in the total energy density of ions and neutrals to the background atmosphere are
\begin{eqnarray}
E_{i}(x,y,t=0) &=& \frac{p_{i}(y)}{\gamma-1} + \frac{|{\bf B}(x,y,t=0)|^{2}}{2} + \frac{p_{i}(y)A_{p}}{\gamma-1}\exp\left(-\frac{(x-x_{0})^{2}+(y-y_{0})^{2}}{w^{2}}\right) = E_{ib} + \Delta E_{i}, \\
E_{i}(x,y,t=0) &=& \frac{p_{i}(y)}{\gamma-1} + \frac{p_{i}(y)A_{p}}{\gamma-1}\exp\left(-\frac{(x-x_{0})^{2}+(y-y_{0})^{2}}{w^{2}}\right) = E_{nb} + \Delta E_{n}.
\end{eqnarray}
\noindent Here the deviations in energies of ions and neutrals from their background values $(E_{i,nb})$ are $\Delta E_{i}$, and $\Delta E_{n}$, respectively. The latter means that the amplitude of the Gaussian pulse determines the energy deviation generated by the perturbation.
%
%
\subsection{Numerical methods}
\label{subsec:numerical_methods}
To solve the two-fluid equations  (\ref{density_ions})-(\ref{divergenceB}) numerically, we employ the JOANNA code \citep{Wojcik_2017}. In all simulations, we set the Courant-Friedrichs-Levy (CFL) number equal to 0.9 and choose the third-order strong stability preserving Runge-Kutta (SSP-RK3) time integrator \citep{2010nmfd.book.....D}. Additionally, we adopt the Harten-Lax-van Leer discontinuities (HLLD) approximate Riemann solver \citep{Miyoshi&Kusano_2005} in combination with a linear reconstruction and the minmod limiter. To control numerically the growth of the solenoidal constraint condition given by equation (\ref{divergenceB}), we use the extended generalized Lagrange multiplier method \citep{Dedner_et_al_2002}. This method is robust in low plasma beta ($\sim 10^{-3}-10^{-2}$) regions, as implied in the solar corona; see, for example, the right panel of Fig. \ref{mag_field}. 

We carry out the simulations in the domain $x\in[-5.12,5.12]$, $y\in[0,60]$, in units of Mm. Here, $y=0$ Mm represents the bottom of the photosphere. In the numerical simulations, we use a uniform grid within the region $(-5.12\le x \le 5.12)$ Mm $\times$ $(0\le y \le 10.24)$ Mm, which is covered by 1024$\times$1024 grid cells, which leads to a spatial resolution of $10$ km. Above this region, within the zone, $(10.24\le y \le 60)$ we implement a stretched grid along the vertical direction $y$; in particular, we divide this region into $64$ grid cells whose size grows with $y$. Such stretched grid helps absorb incoming signals and minimizes reflections from the top boundary \citep{Srivastava_et_al_2017, Kuzma_et_al_2017b}. We impose outflow boundary conditions at the side edges specified by $x=-5.12$ Mm and $x=5.12$ Mm. We set all the plasma quantities to their equilibrium values at the bottom and top boundaries delimited by $y=0$ Mm and $y=60$ Mm.
%
\section{Results of numerical simulations}
\label{sec:results_of_numerical_simulations_analysis}
%
\subsection{Parameters}
\label{subsec:parameters_analyzed}
The equilibrium magnetic field, defined by equations (\ref{Bx_field}) and (\ref{By_field}) and displayed in Fig. \ref{mag_field}, represents a magnetic null point in a potential arcade that extends up to the solar corona. In the context of the solar atmosphere, the magnetic arcades typically refer to active region structures; however, in the quiet Sun, an arcade may represent the conditions corresponding to a bipolar region of a network magnetic field. The maximum magnetic field  strength takes place at the base of the magnetic arcade and it reaches about $50$ G, see, for example, \cite{Choe_et_al_1996} and \cite{Bagwell_et_al_2020}. Regarding the null points on the Sun, some estimations indicate that their number is roughly one per supergranular cell above the height $1.5$ Mm \citep[see, e.g.,][]{Close_et_al_2004, Regnier_et_al_2008}, and likely more below that height. However, numerical simulations and extrapolations of photospheric magnetic fields predict ubiquitous null points in the solar chromosphere and the corona \citep{1997JGR...102..231G, 2005LRSP....2....7L}. Based on the above references, from all the possible combinations of the magnetic field strength and the location of the null point in the solar atmosphere, we select two illustrative ones, which are labeled as Run\#1 and Run\#2, and displayed in Table \ref{parameters}. These runs represent two magnetic field strengths in the upper range of the typical values of the arcades; however, the magnitude of the fields is about $100$ G, which is close to the maximum magnetic field strength at the base of the magnetic arcade. The chosen parameters indicate two locations of the null points at the chromosphere ($Y=1.6$ Mm) and at the corona ($Y=2.2$ Mm). In the following subsections, we describe in detail the results corresponding to Run\#1 and some representative results for Run\#2. 
\begin{table}
    \centering
    \begin{tabular}{|c|c|c|c|c|c|}
    \hline
    & $h$ (Mm) & $Y$ (Mm) & $b$ (G) & $B_{0}$ (G) & $y_{0}$ (Mm) \\
    \hline    
    Run\#1  & 0.3 & 1.6 & 10 & 401 & 1.6 \\
    \hline
    Run\#2  & 0.3 & 2.2 & 10 & 684 & 2.2 \\
      \hline
    \end{tabular}
    \caption{Parameters used for Run\#1 and Run\#2.}
    \label{parameters}
\end{table}
%
\subsection{Morphology of the jets}
\label{subsubsec:results_Run1_Run2}
\begin{figure*}
\centering
\centerline{\Large \bf   
      \hspace{0.433\textwidth}  \color{black}{\large{Run\#1}}
         \hfill}
\vspace{0.3cm}           
\includegraphics[width=5.5cm,height=5.5cm]{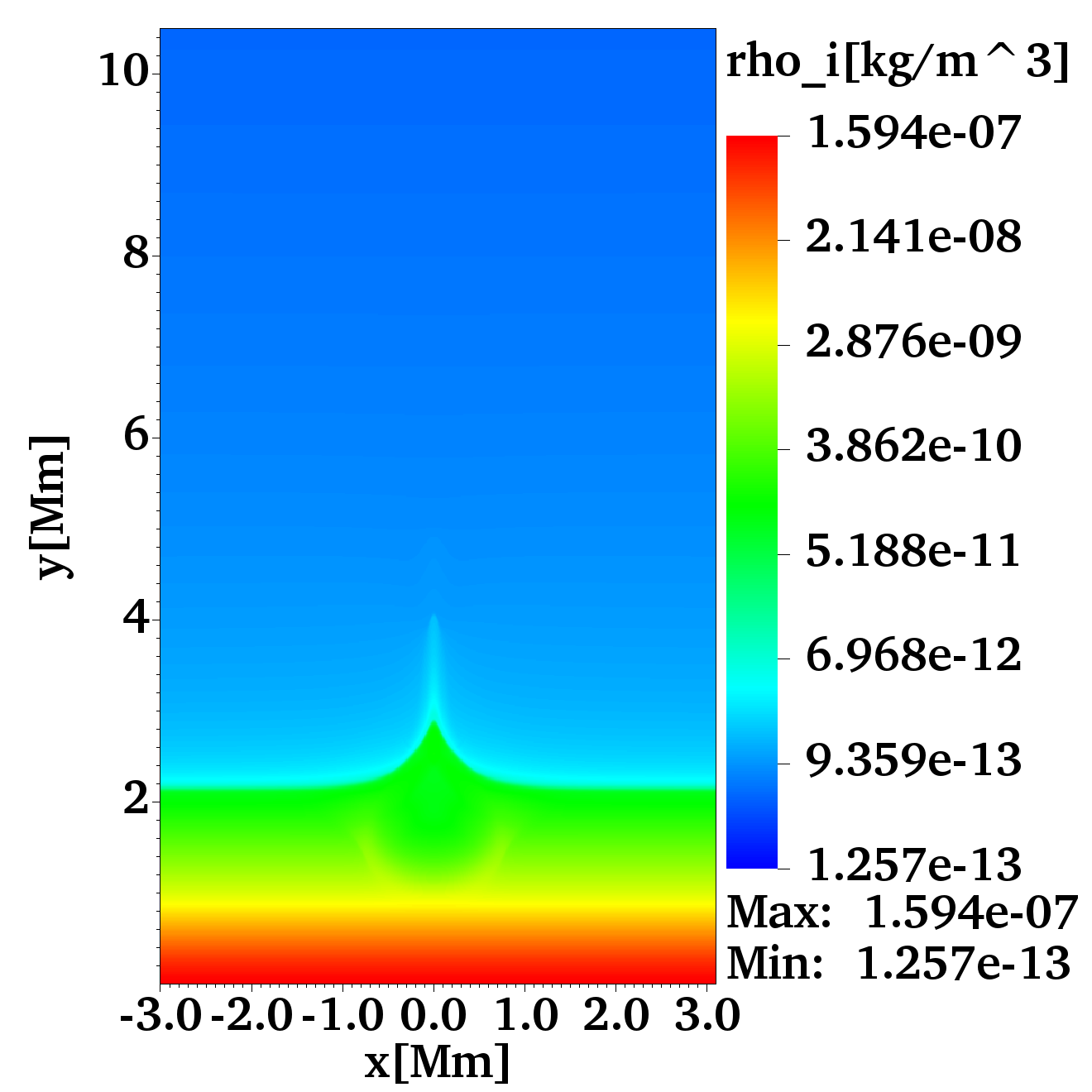}
\includegraphics[width=5.5cm,height=5.5cm]{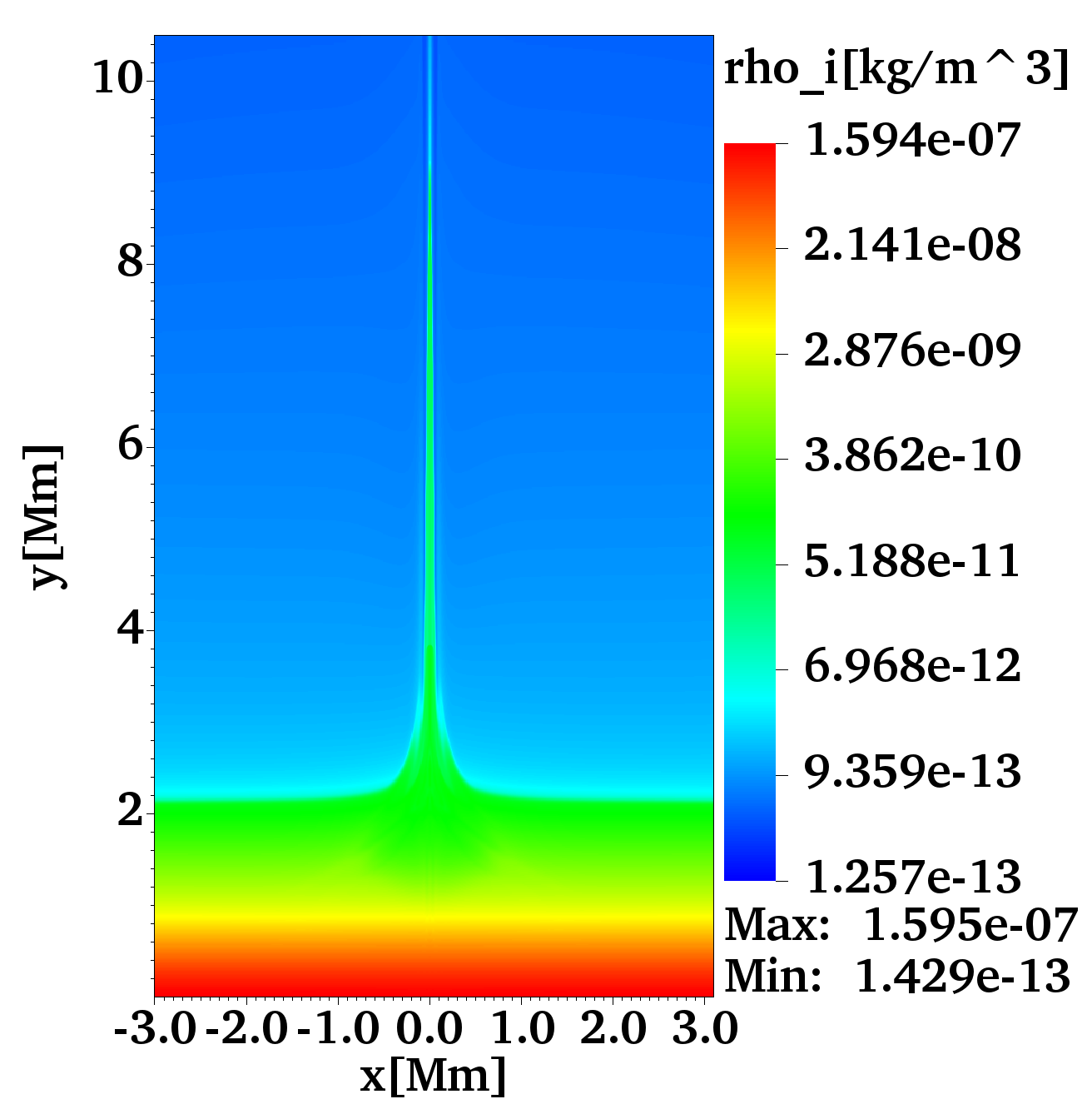}
\includegraphics[width=5.5cm,height=5.5cm]{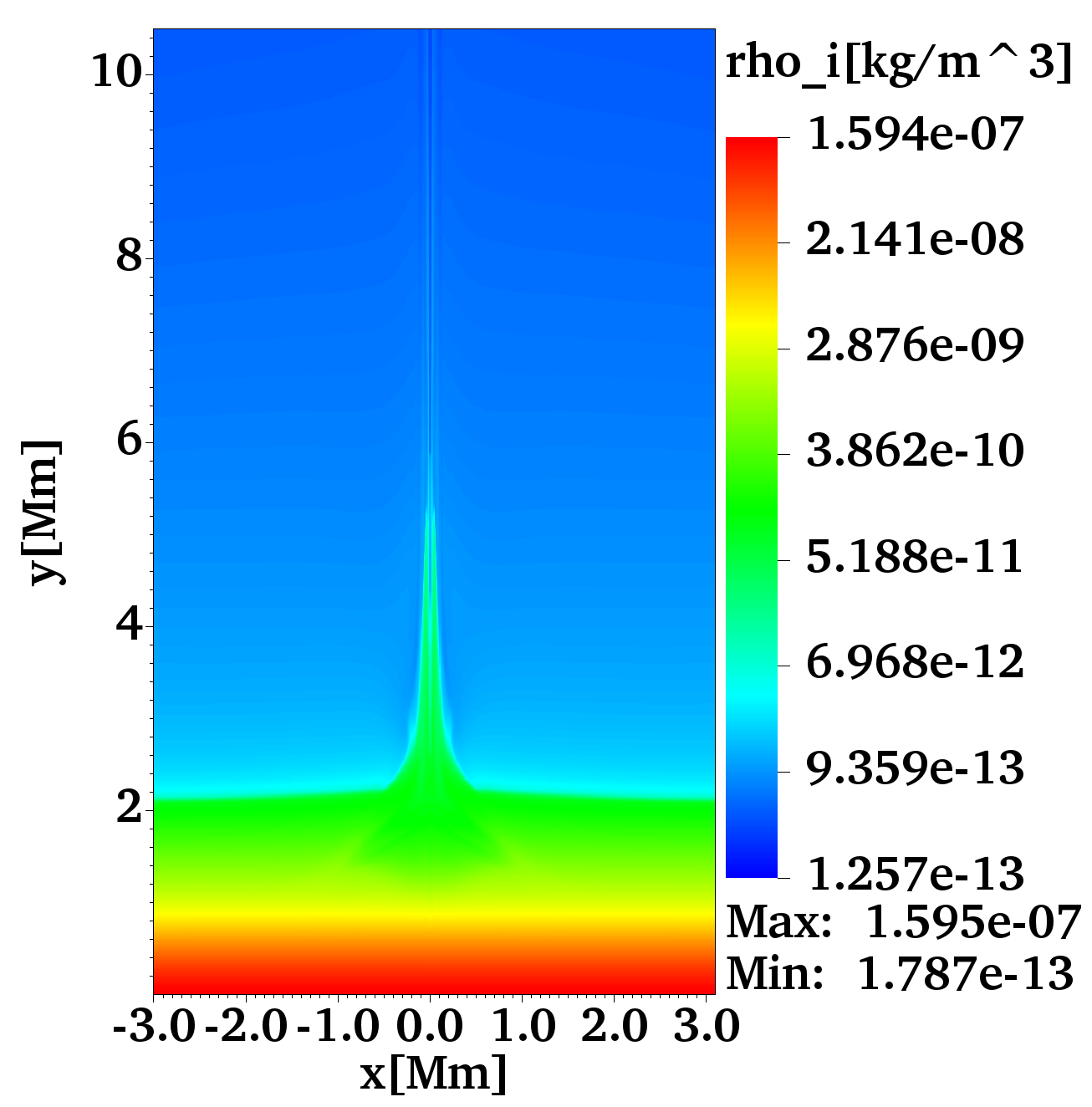}
\includegraphics[width=5.5cm,height=5.5cm]{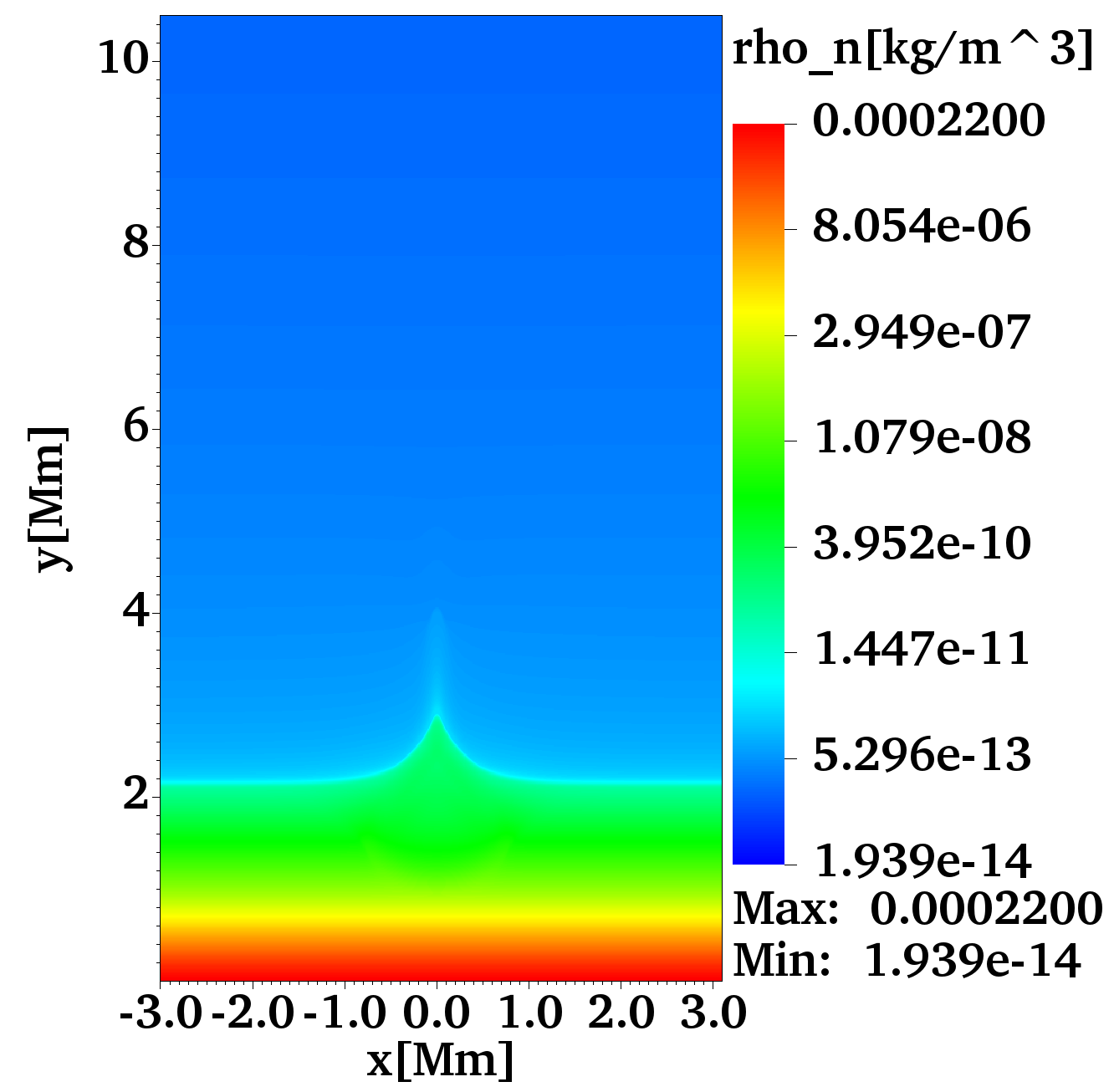}
\includegraphics[width=5.5cm,height=5.5cm]{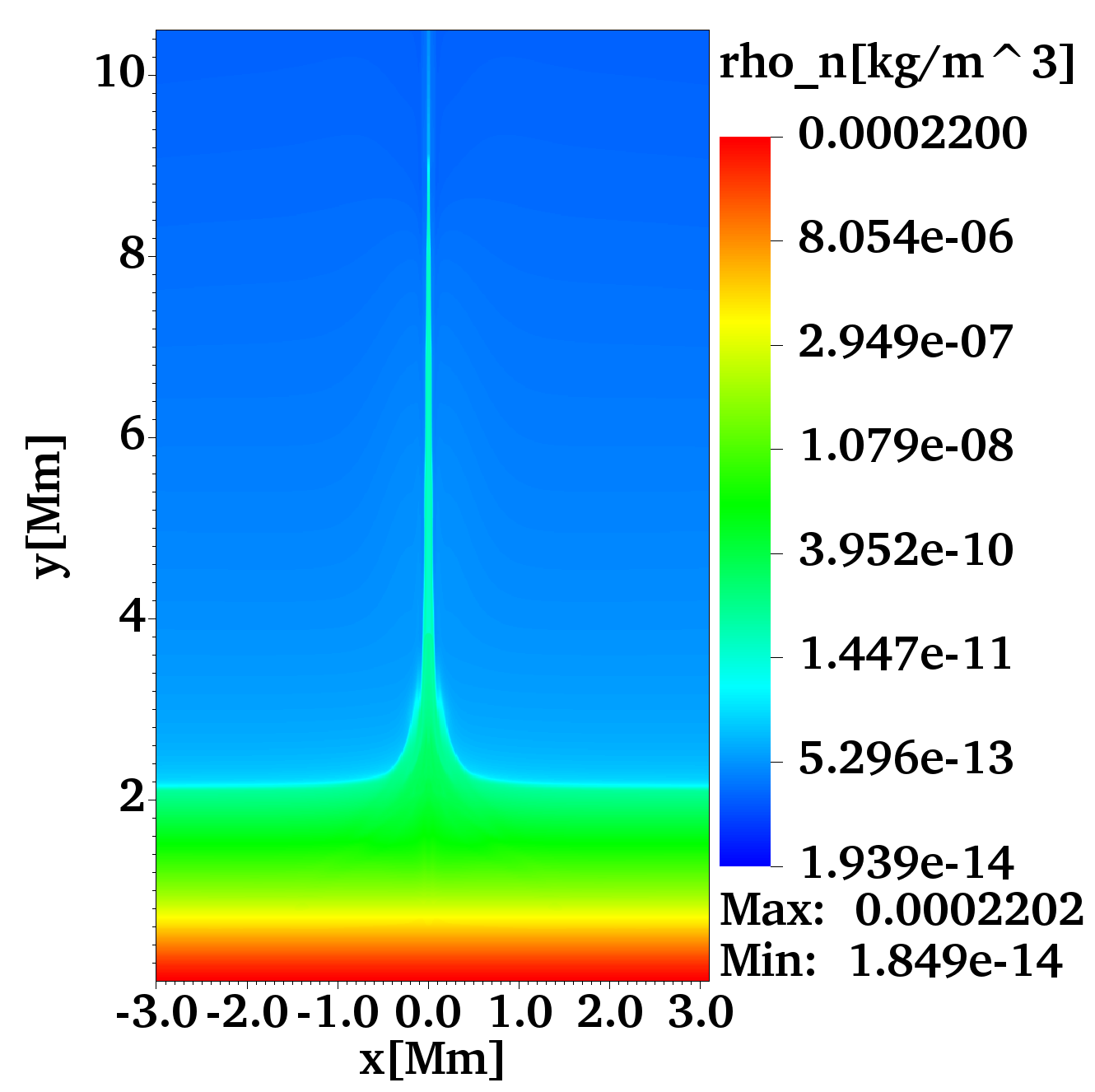}
\includegraphics[width=5.5cm,height=5.5cm]{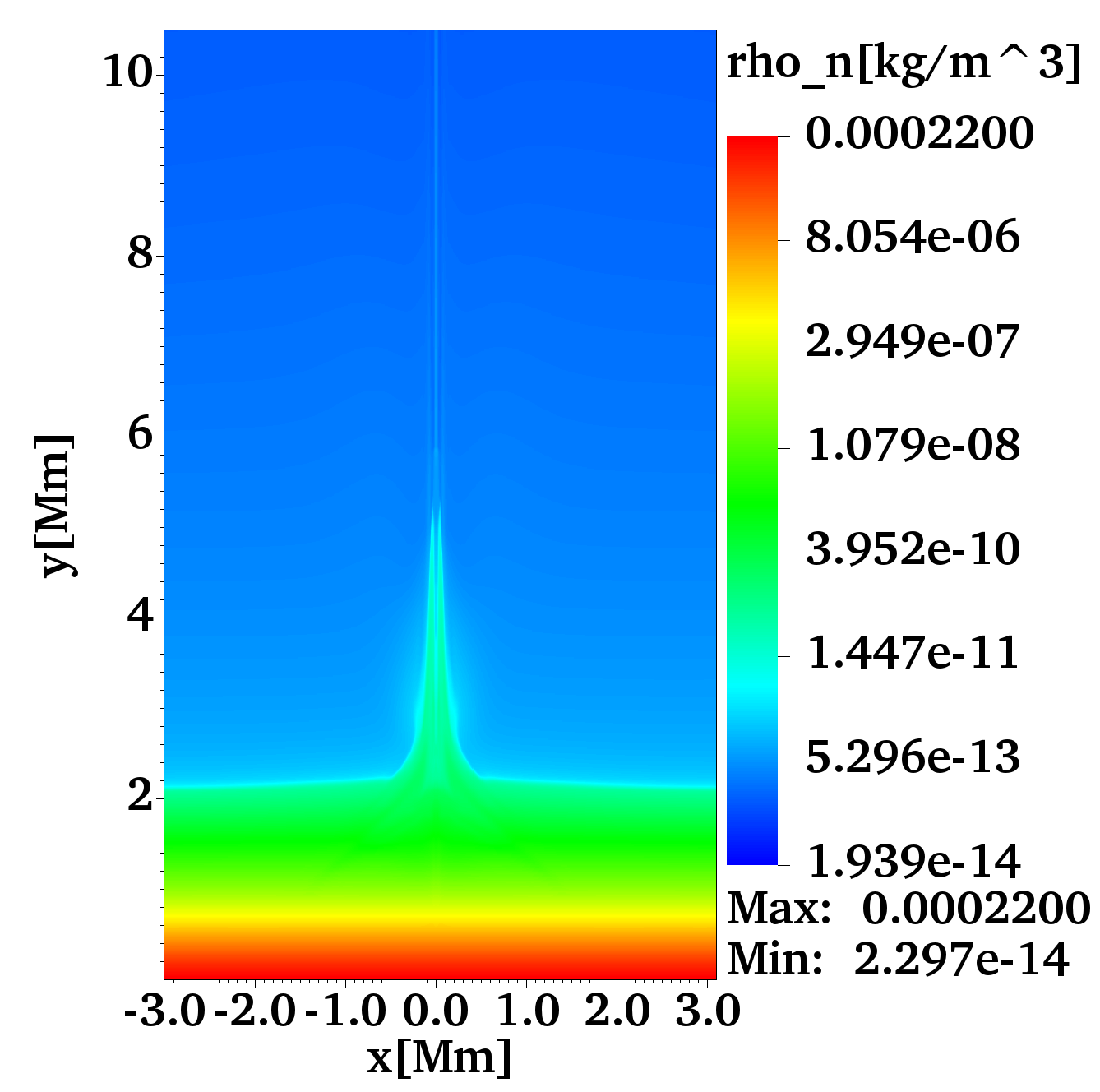}
\caption{Temporal evolution of log($\varrho_{i}(x,y)$) (top), and log($\varrho_{n}(x,y)$) (bottom) corresponding to Run\#1 at $t=30$ s, $300$ s, and $600$ s from left to right.}
\label{fig:density_maps_of_ions_run1_with_interactions}
\end{figure*}

In Fig. \ref{fig:density_maps_of_ions_run1_with_interactions}, we show the mass density of ions, $\varrho_{i}$, (top panels) and neutrals $\varrho_{n}$ (bottom panels), expressed in kg m$^{-3}$ and corresponding to  Run\#1. In the top-left, we show a snapshot of $\varrho_{i}$ at $t=30$ s. Here, we identify the formation of a collimated plasma structure consisting of two parts: (i) the bottom part that mimics an inverted shaped-V jet of $\varrho_{i}\sim10^{-11}$ kg m$^{-3}$, which is larger than the background $\varrho_{i}\sim10^{-13}$ kg m$^{-3}$ and it reaches a height of approximately 3 Mm; (ii) the top part, which is a low, dense plasma of ions ($\varrho_{i}\sim10^{-12}$ kg m$^{-3}$) which propagates upwards up to about $4$ Mm. The physical mechanism of the jet formation is the following: the pressure pulse initially launched from the solar chromosphere ($y=1.6$ Mm), propagates upwards, grows in its amplitude due to the decrease of density with height to conserve momentum $\varrho_{i} v_{iy}$ $c_{i}$, and steepens into a shock. This shock propagates upward into the corona, but chromospheric plasma lags behind the shock front, which rises to about $y=4$ Mm. The chromospheric material is lifted so that the plasma's rarefaction resibling behind the shock front, leading to low pressure. As a result, the pressure gradient force works against the gravity and pushes the chromospheric material to penetrate the solar corona \citep[see, e.g.,][]{Muraswki&Zaqarashvilli_2010}. The next snapshot (top-middle), at $t=300$ s, shows that the chromospheric plasma propagates upwards and becomes narrow. This behavior is consistent with the fact that ions follow the magnetic field lines, which are vertical at the corona, as shown in Fig. \ref{mag_field}. In the top-right, the mass density of ions at $t=600$ s illustrates that the plasma flows downwards because of the gravity action. At this time, a double structure is visible at the top of the jet. This double structure appears when the plasma falls back at the central part of the jet and is forced to move upwards again at the jet boundaries due to the next shock. At the bottom, we display the mass densities of neutrals for the same three instants of times ($t=30, 300, 600$ s) that resemble similar distribution to $\varrho_{i}$. However, we can indicate some differences. For example, at $t=30$ s, the inverted-Y shape of the jet is slightly smaller ($\sim 3$ Mm) than for ions. At $t=600$ s, the distribution of mass density of neutrals is wider ($\sim 0.5$ mm) than the mass density of ions.

\begin{figure*}
\centering
\centerline{\Large \bf   
      \hspace{0.433\textwidth}  \color{black}{\large{Run\#2}}
         \hfill}
\vspace{0.3cm}         
\includegraphics[width=5.5cm,height=5.5cm]{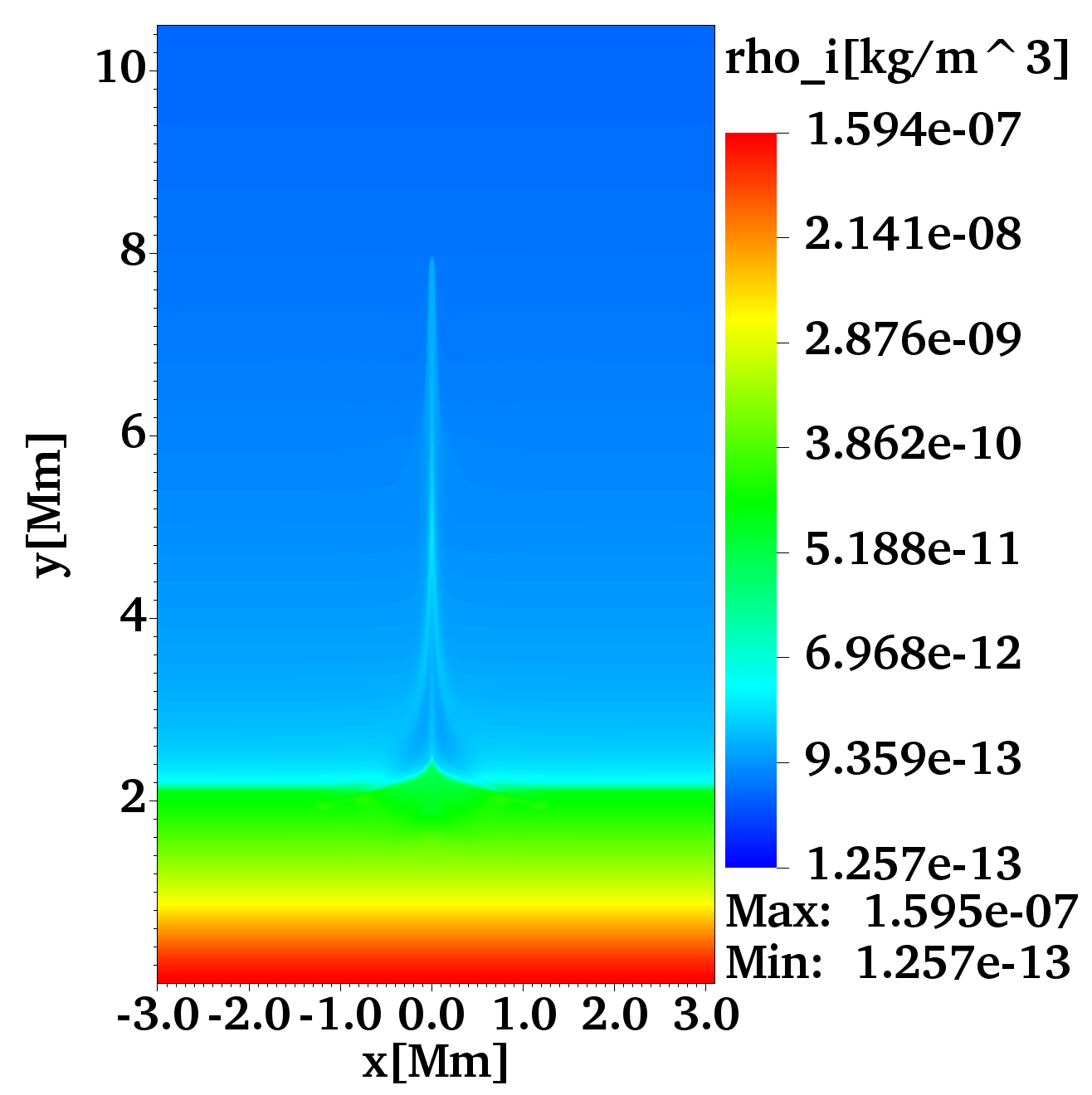}
\includegraphics[width=5.5cm,height=5.5cm]{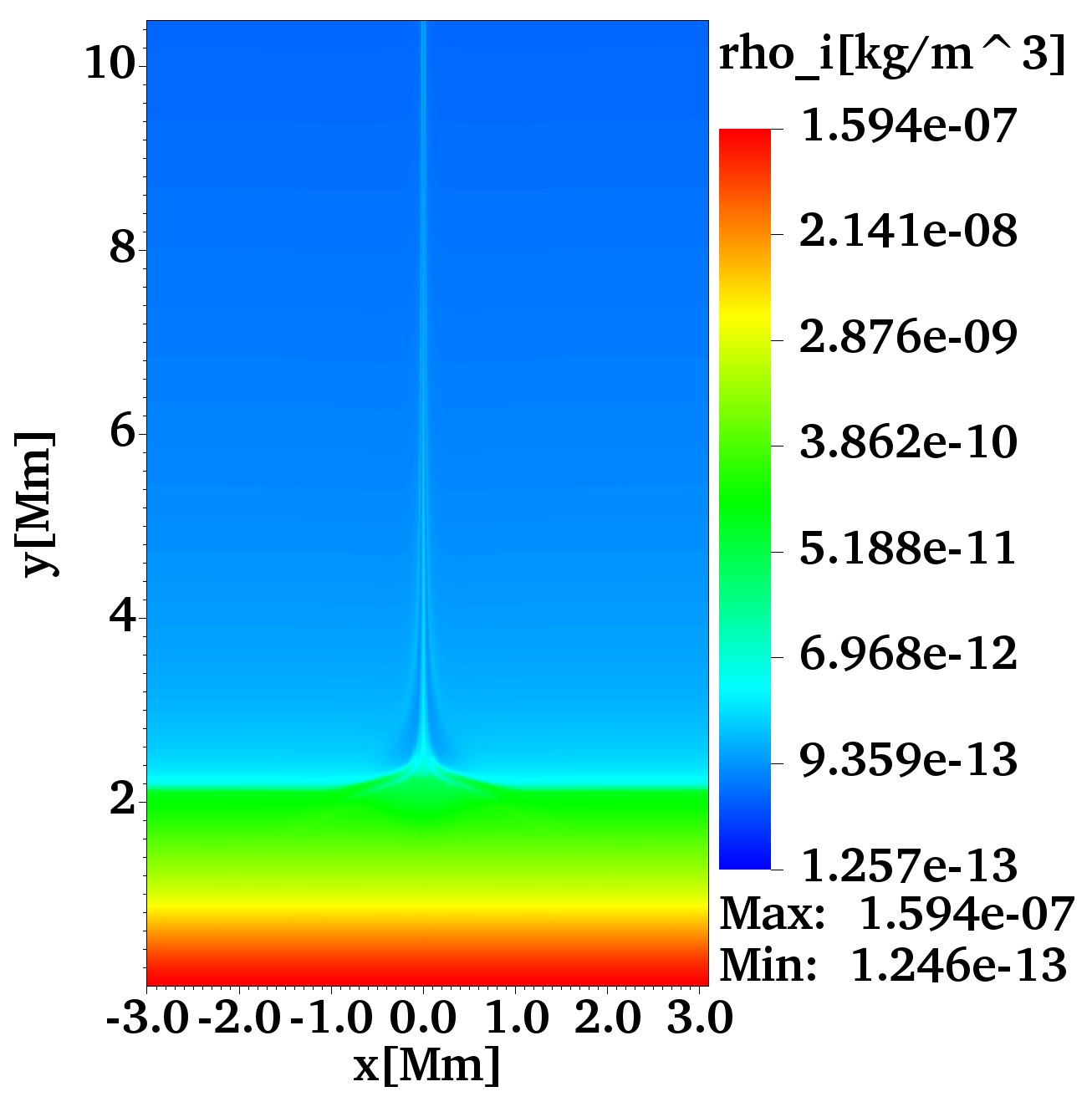}
\includegraphics[width=5.5cm,height=5.5cm]{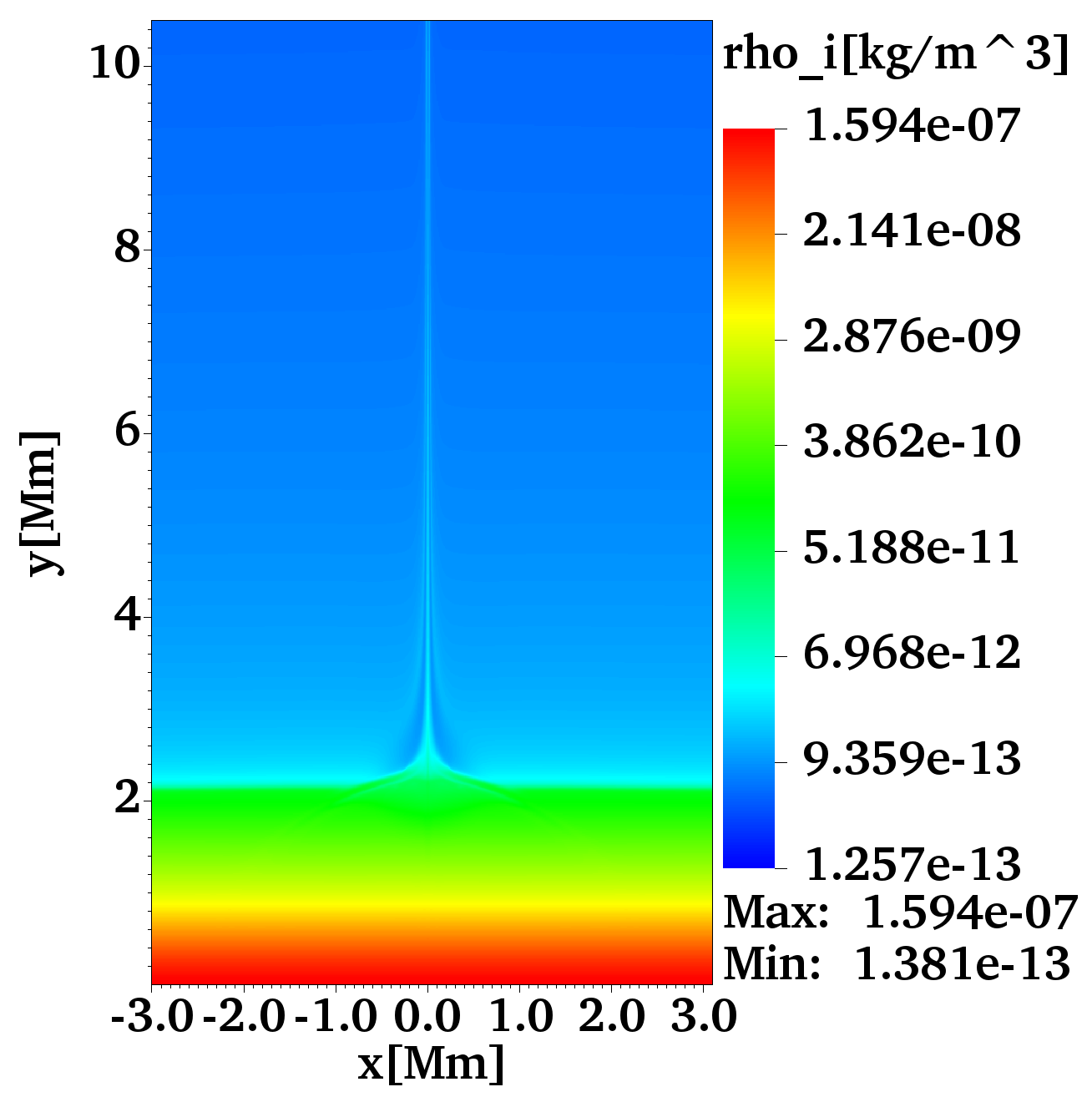}
\caption{Temporal evolution of log($\varrho_{i}(x,y)$) corresponding to Run\#2 at $t=30$ s, $300$ s, and $600$ s.}
\label{fig:density_maps_of_ions_run2_with_interactions}
\end{figure*}

In Fig. \ref{fig:density_maps_of_ions_run2_with_interactions} we display snapshots of $\varrho_{i}$ at three instants of time $t=30, 300, 600$ s for Run\#2. In this scenario, $\varrho_{n}$ resembles $\varrho_{i}$, so we decided not to show them. On the left, at $t=30$ s, no collimated plasma jet formation is discernible, as in Run\#1. Instead, we see a small structure of about $10^{-11}$ kg m$^{-3}$ that practically remains at its initial position, i.e., at $y=2.2$ Mm. At $t=30$ s, we can also see that the neutrals with a smaller value of $\varrho_{n}\approx10^{-12}$ kg m$^{-3}$ reach a height of approximately $8$ Mm. However, this structure does not mimic any characteristic of a jet. It is rather the response of the low, dense background atmosphere with $\varrho_{n}\sim10^{-13}$ kg m$^{-13}$ to the initial pressure pulses. That is, in this scenario, the location of the pressure pulse is over the TR, at $y=2.2$ Mm, and it is present in a low-density region with $\varrho_{n}\approx 10^{-13}$ kg m$^{-3}$; these conditions are not suitable for the formation of any jet as it takes place in Run\#1. At $t=300$ s (middle), the less dense in $\varrho_{n}$ structure becomes elongated, while the denser part with $\varrho_{n}\sim10^{-11}$ kg m$^{-3}$ remains just over $y=2.2$ Mm. Finally, on the right, at $t=600$ s, the mass density of neutrals is similar to the time $t=300$ s. 

Although we impose outflow boundary conditions at the sides edges $x=\pm 5.12$ Mm, the system's total mass was not significantly lost. We estimate the relative percentage of mass loss between the initial and final states of the simulations, given by $\Delta\varrho_{rel} = \frac{\Delta\varrho_{max-tf}}{\varrho_{max-bg}}*100\%$, where $\Delta\varrho_{max-tf}$ is the maximum difference of the mass density between the initial and final states, and $\varrho_{max-bg}$ is the maximum value of the background density at the final state of the simulations. In the particular case of Run\#1, the relative percentage of the loss mass of ions was about 0.1\%, while it was about 0.01\% for neutrals. Therefore, the relative total mass loss was about 0.11\% for Run\#1. For Run\#2, the relative percentage of mass loss for ions was 0.01\%, while for neutrals, it was approximately 0.001\%; then, the relative total mass loss for Run\#2 was about 0.011\%. So practically, the total mass was conserved in these two simulations.
%
\subsection{Time evolution of jets}
\label{subsec:time-series_time_distance_plots}
\begin{figure*}
 \centerline{\Large \bf   
      \hspace{0.245\textwidth}  \color{black}{\Large{Run\#1}}
       \hspace{0.36\textwidth}  \color{black}{\Large{Run\#2}}
          \hfill}
        \includegraphics[width=8.0cm,height=5.5cm]{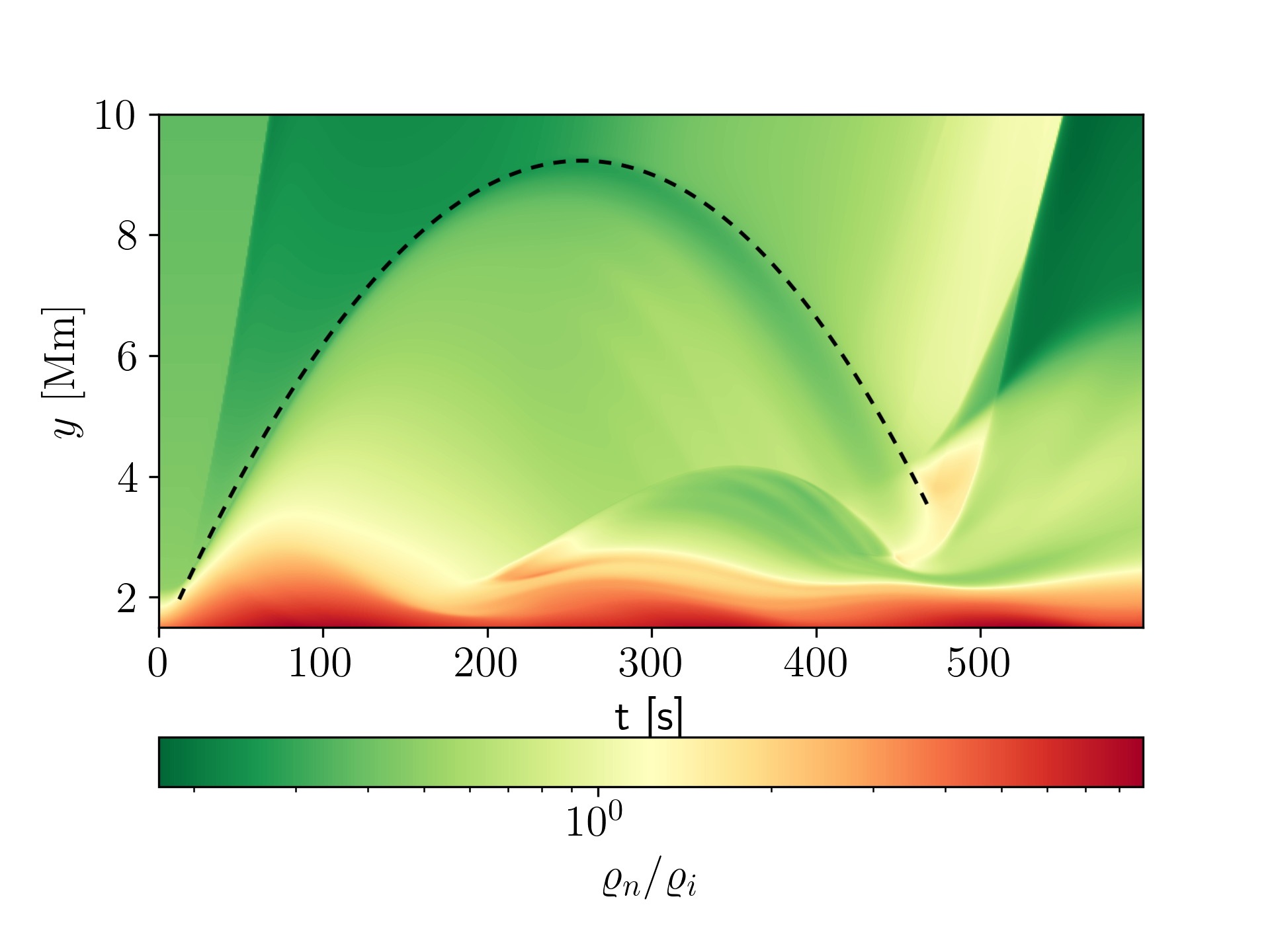}
       \includegraphics[width=8.0cm,height=5.5cm]{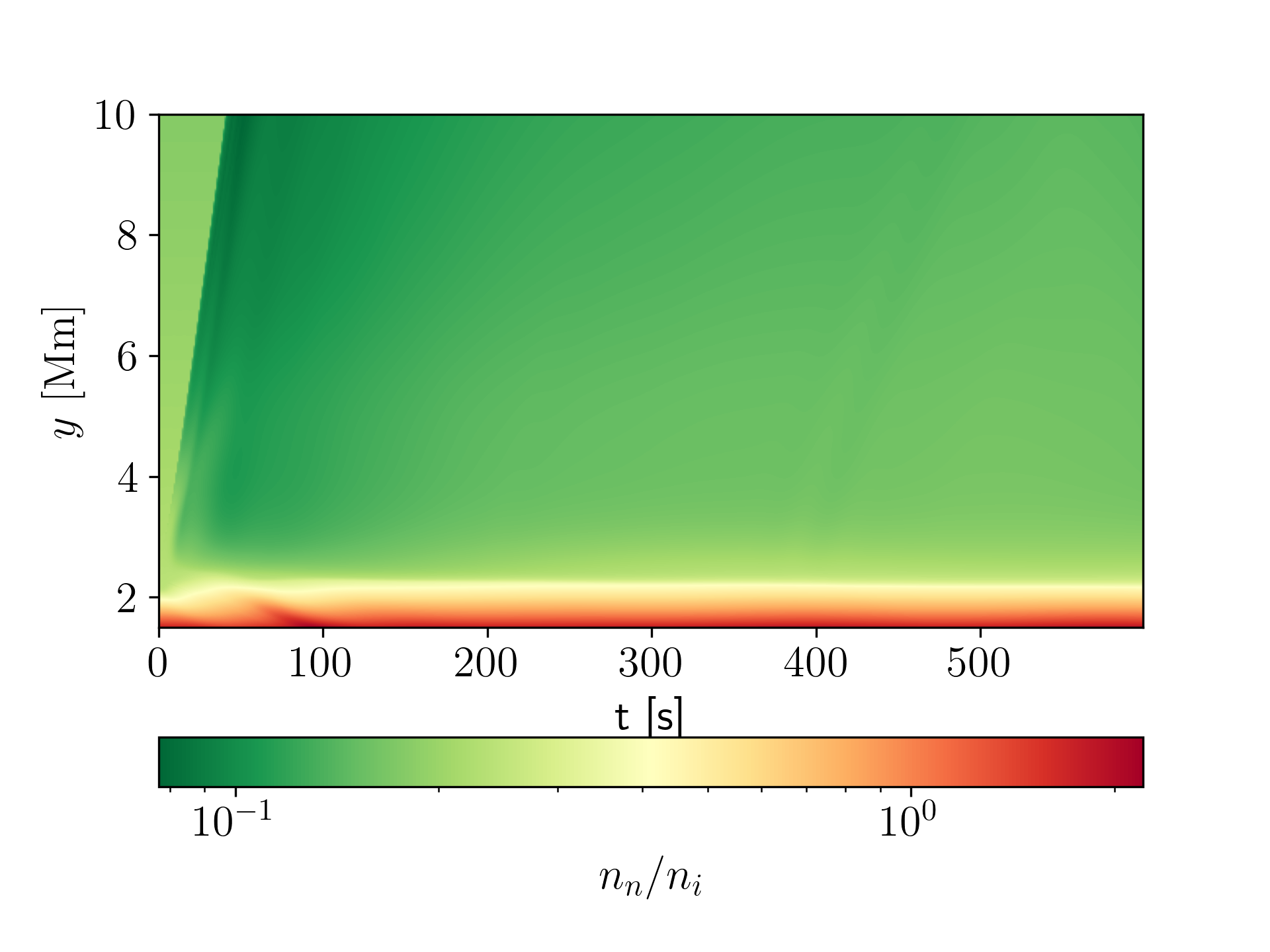}
        \includegraphics[width=8.0cm,height=5.5cm]{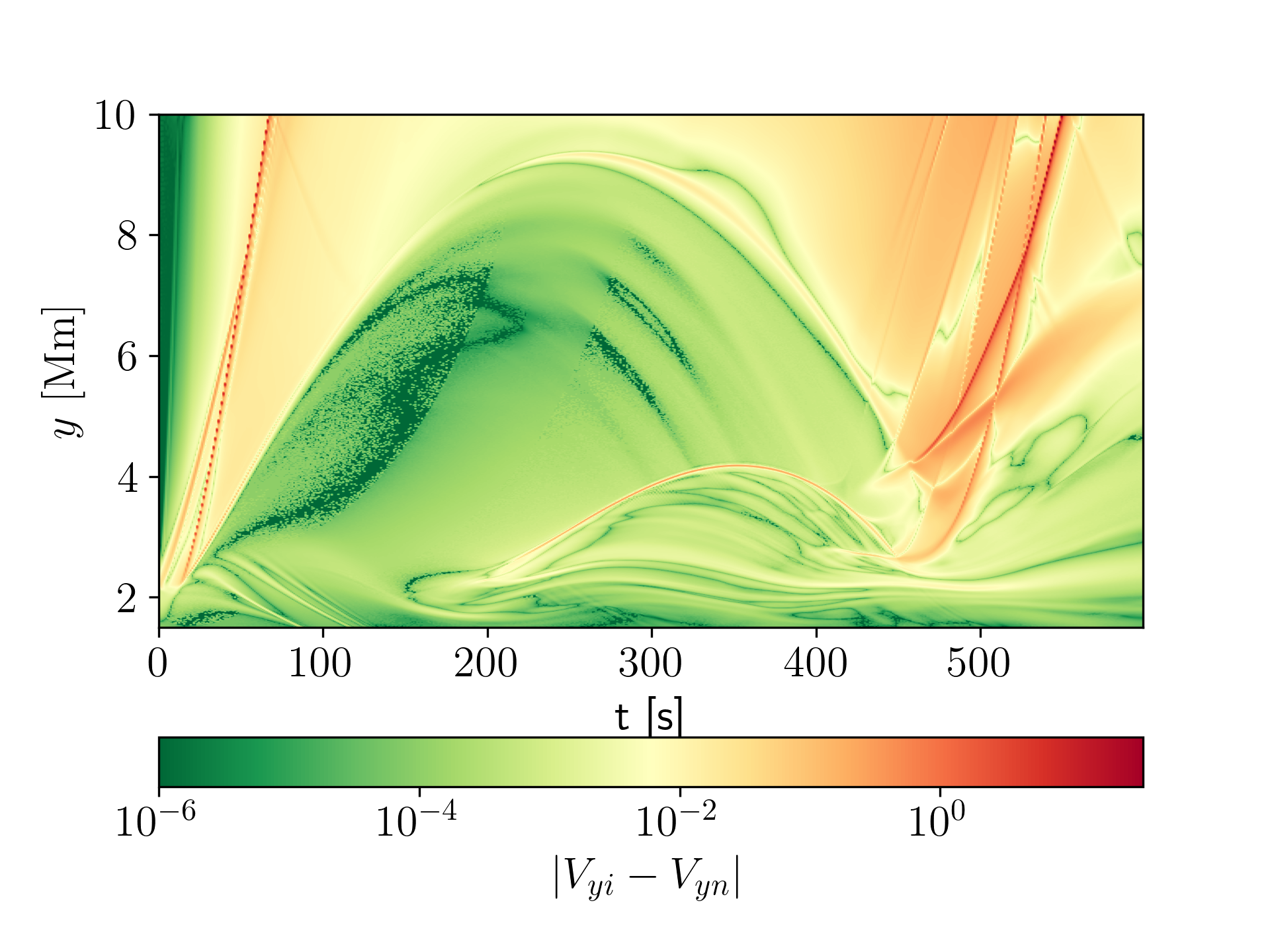}
      \includegraphics[width=8.0cm,height=5.5cm]{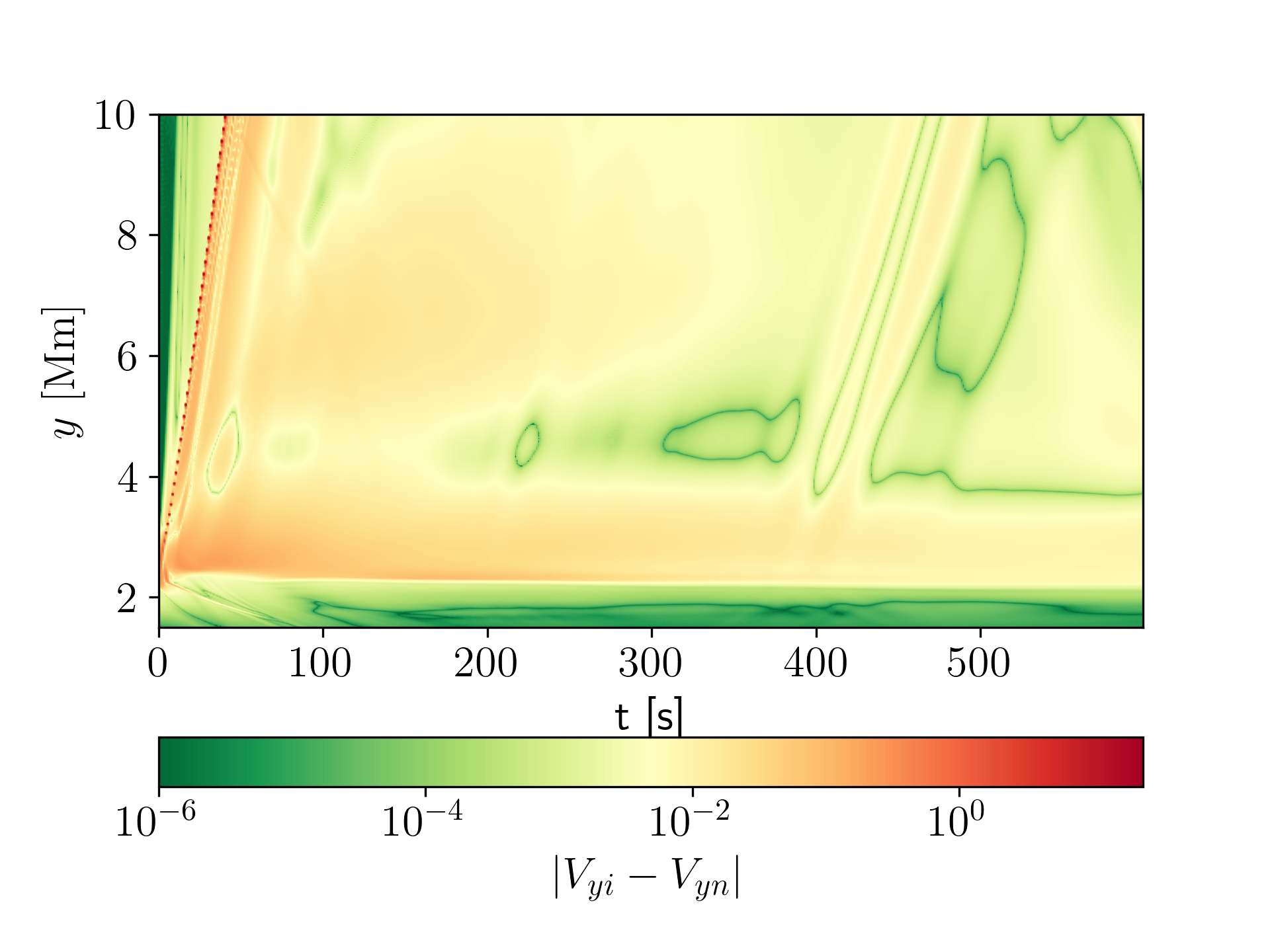}
\caption{Time-distance plots for the $\log(If(x=0,y))$ (left) and $\log(|V_{y_{i}}(x=0,y)-V_{y_{n}}(x=0,y)|$ for Run\#1 (left panels) and for Run\#2 (right panels).}
\label{fig:Distance-time_diagrams_If_vel_drift}
\end{figure*}

To show the evolution of the ions and neutrals, and identify their potential differences of Fig. \ref{fig:density_maps_of_ions_run1_with_interactions}, we plot the temporal evolution of the ionization fraction $\varrho_{n}/\varrho_{i}$ and the absolute value of vertical velocity difference $|V_{yi}-V_{yn}|$ along vertical cuts in $y$, in Fig. \ref{fig:Distance-time_diagrams_If_vel_drift} for Run\#1 (left panels) and Run\#2 (right panels). In the top-left panel, we show the ionization fraction for Run\#1, from which we infer that the temporal evolution of ions and neutrals differs one from another since, at lower altitudes ($y\sim2$ Mm), neutrals are more predominant than ions, while at higher altitudes ($y>2.1$ Mm) the ions are more abundant in the evolution. In particular, we identify with the dashed line that ions and neutrals follow a parabolic path, i.e., there is material moving upwards that later falls back. This parabolic path is quasi-periodic since the next shocks are coming up from the chromosphere (see below). In the top-right, for Run\#2, we see that the ionization fraction does not vary in the time since neutrals are predominant at lower altitudes, while ions consist of the abundant component at higher altitudes. Meanwhile, in the bottom-left panel for Run\#1, the logarithm of the absolute difference between vertical components of ion and neutral 
velocities, $\log|V_{yi}-V_{yn}|$, exhibits some differences between the kinematics of ions and neutrals. We note that the difference between these vertical velocities attains a minimum along the parabolic path, which means that the vertical velocity of ions is close to the vertical velocity of neutrals. It means that the jet carries ions and neutrals, which are lifted up 
as a part of the chromospheric material that lags behind the shock. In contrast, in the surroundings of the jet, in the solar corona, the vertical velocity of ions is greater than that of neutrals, as it is expected for a predominant region of ions. Finally, in the bottom-right for Run\#2, the vertical velocity difference is close to zero at lower altitudes, while it is slightly greater at higher altitudes, implying that ions are predominant in the solar corona.

\begin{figure*}
\includegraphics[width=7.5cm,height=5.5cm]{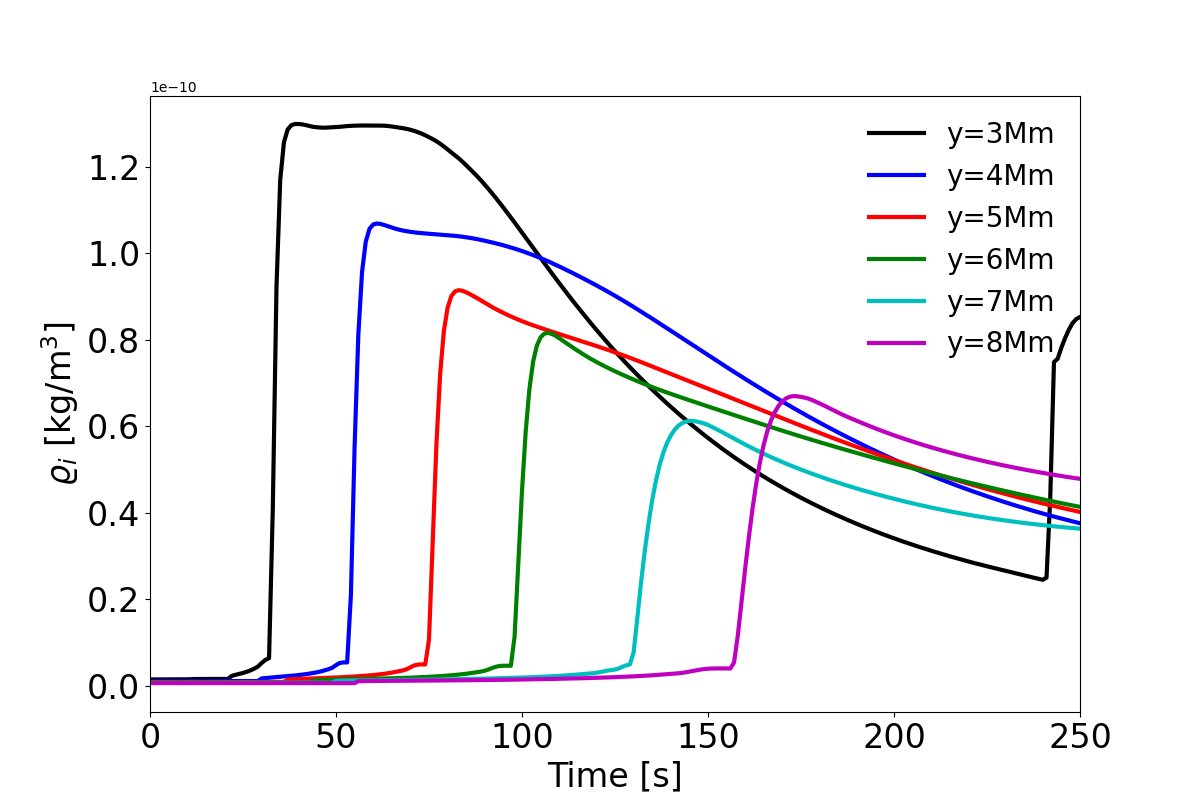}
\caption{Time series of the mass density of ions, $\varrho_{i}$, collected at the points given by $x=0,y=3,4,5,6,7,8$ Mm for Run\#1.}
\label{fig:evolution_waves_Run1}
\end{figure*}

We can identify the global behavior of the mass density of ions for Run\#1 by the time signatures collected for $x=0$ Mm at different detection points along the vertical direction $y$, as shown in Fig. \ref{fig:evolution_waves_Run1}. This plot essentially reveals the shock's presence, represented by the steep gradient in the mass density. This shock propagates and passes $y=3$ Mm reaching the level $y=8$ Mm. In this figure, it is clear that the shock is strong just after forming, i.e., at $y=3$ Mm, while it becomes weak when it reaches upper heights. For example, at $y=6$ Mm (green line), the steep gradient of ion mass density ($\varrho_{i}\sim 10^{-11}$ kg m$^{-3}$) at about $100$ s is two orders of magnitude higher than in the ambient coronal plasma ($\varrho_{i}\sim10^{-13}$ kg m$^{-3}$).

\begin{figure*}
         \centerline{\Large \bf   
      \hspace{0.26\textwidth}  \color{black}{\Large{Run\#1 for a straight magnetic field}}
         \hfill}
         \includegraphics[width=8.0cm,height=5.5cm]{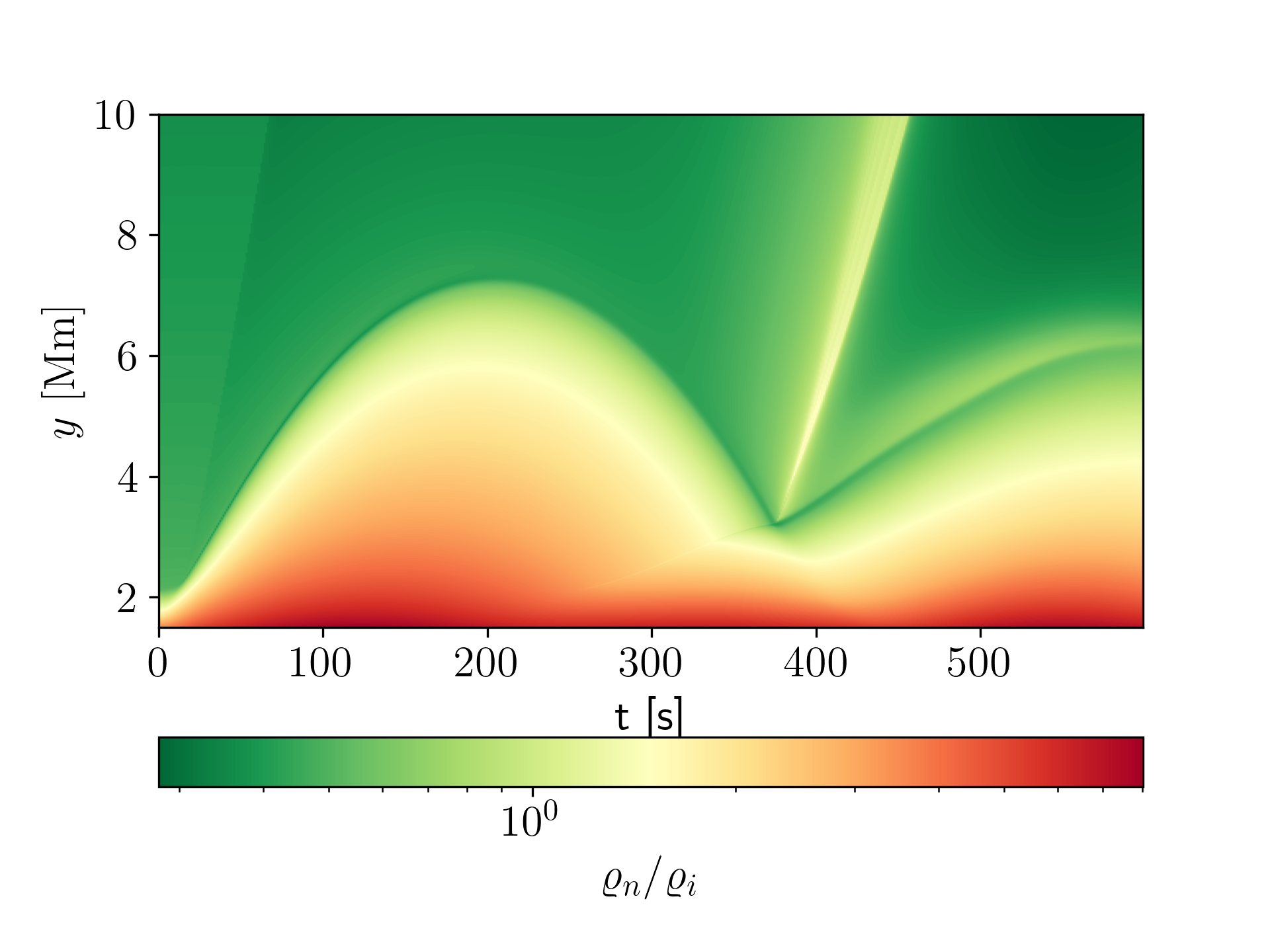}
      \includegraphics[width=8.0cm,height=5.5cm]{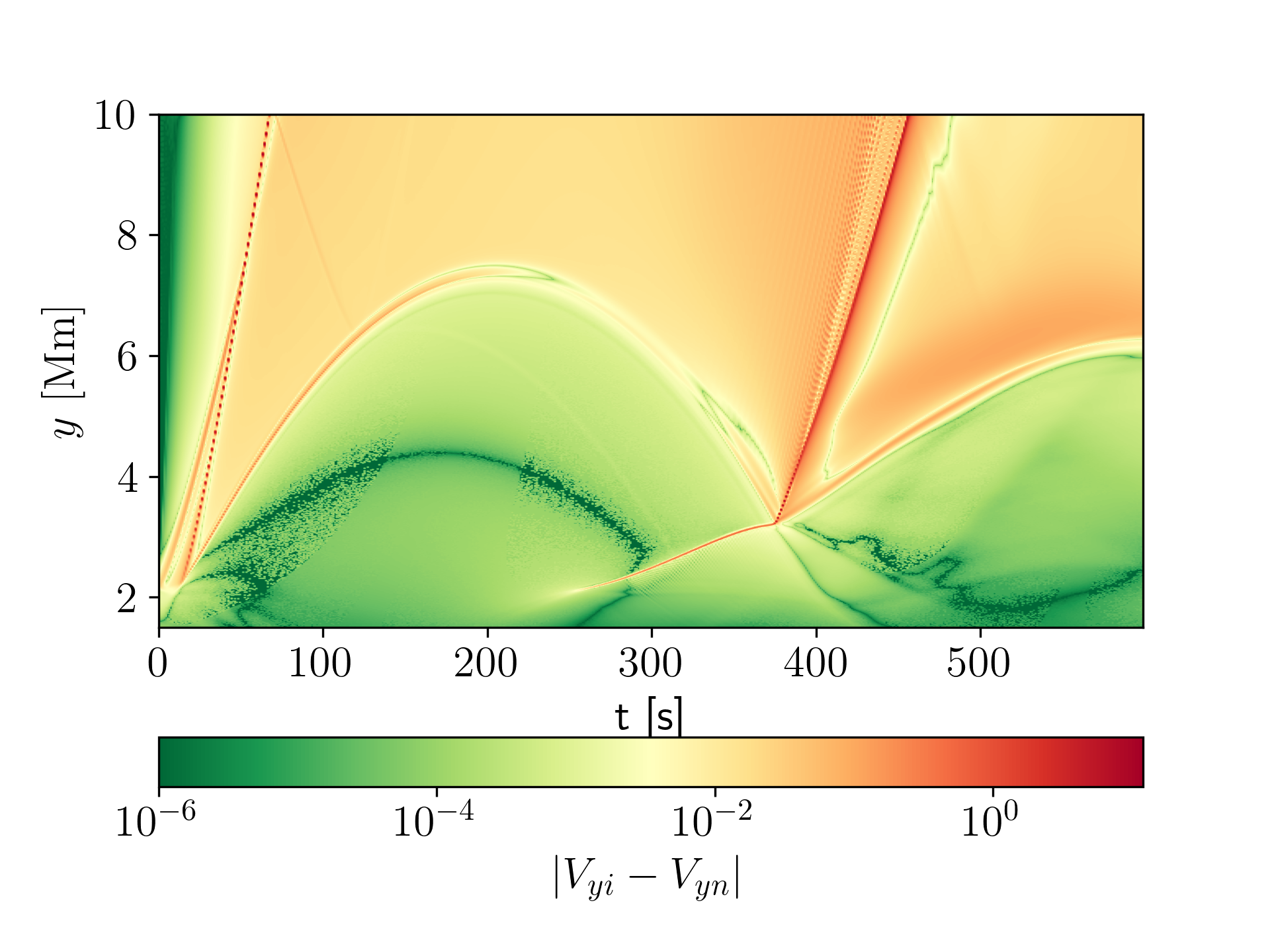}
    \caption{Time-distance plots for the $\log(If(x=0,y))$ (left) and $\log(|V_{y_{i}}(x=0,y)-V_{y_{n}}(x=0,y)|$ (right) corresponding to Run\#1 with a straight magnetic field configuration.}
\label{fig:Distance-time_diagrams_Run1_with_interactions_straight_field}
\end{figure*}

It is interesting to check whether the behavior of the jets is related to the magnetic null point and the potential arcade. Therefore, we perform an additional simulation considering a constant and straight magnetic field with its magnitude equal to $10$ G for the exact parameters of the pressure pulses corresponding to Run\#1, i.e., $A_{p}=8.5$ and $y_{0}=1.6$ Mm. For this simulation, we only show the time-distance plots of the logarithm of the ionization fraction,  $\varrho_{n}/\varrho_{i}$, and the absolute value of the vertical velocity difference $|V_{yi}-V_{yn}|$, which are displayed in Fig.  \ref{fig:Distance-time_diagrams_Run1_with_interactions_straight_field}. For instance, on the left, the ionization fraction shows that ions and neutrals are distributed more homogeneously over the parabolic path than the results shown in the top-left panel of Fig.  \ref{fig:Distance-time_diagrams_If_vel_drift}. This behavior is related to the magnetic field configuration since, in the case of the null point in the potential arcade, the ions are not guided by the magnetic field from the start of the simulation; instead, they propagate over a free-magnetic region and then feel the presence of the magnetic field. On the other hand, for a vertical straight field, the ions are guided from the start of the simulation, so the pressure pulses propagate and then steepen into a shock that lifts the chromospheric material, which contains a large amount of neutrals. 
However, in the background density of the solar corona, ions are more abundant than neutrals. The vertical velocity difference shown in the right panel of Fig.  \ref{fig:Distance-time_diagrams_Run1_with_interactions_straight_field}, behaves consistently as the ionization fraction does. We can see that the vertical velocity difference is near zero over the distribution that defines the parabolic path, while in the solar corona, the vertical velocity of ions is greater than neutrals. Then, we can specifically distinguish the following difference between the scenario with the magnetic null point in the potential arcade and the vertical straight field: (i) the maximum height reached by the mass densities of ions and neutrals of $\sim 9$ Mm is greater than the reached in the vertical magnetic field case $\sim7$ Mm; (ii) the parabolic path of the mass density of ions and neutrals for the case of the null point seem to have a greater period of oscillations $\sim 450$ s than  $\sim 350$ s shown in the vertical magnetic field case; (iii) the ions and neutrals behave more homogeneously over the parabolic path in the case of a vertical straight field than in the case of the null point in the potential arcade. 

\subsection{Time series of ions and neutrals at the null point and the cut-off period}
\label{subsec:dynamics_morphology_around_null_point}

\begin{figure*}
     \includegraphics[width=8.0cm,height=5.5cm]{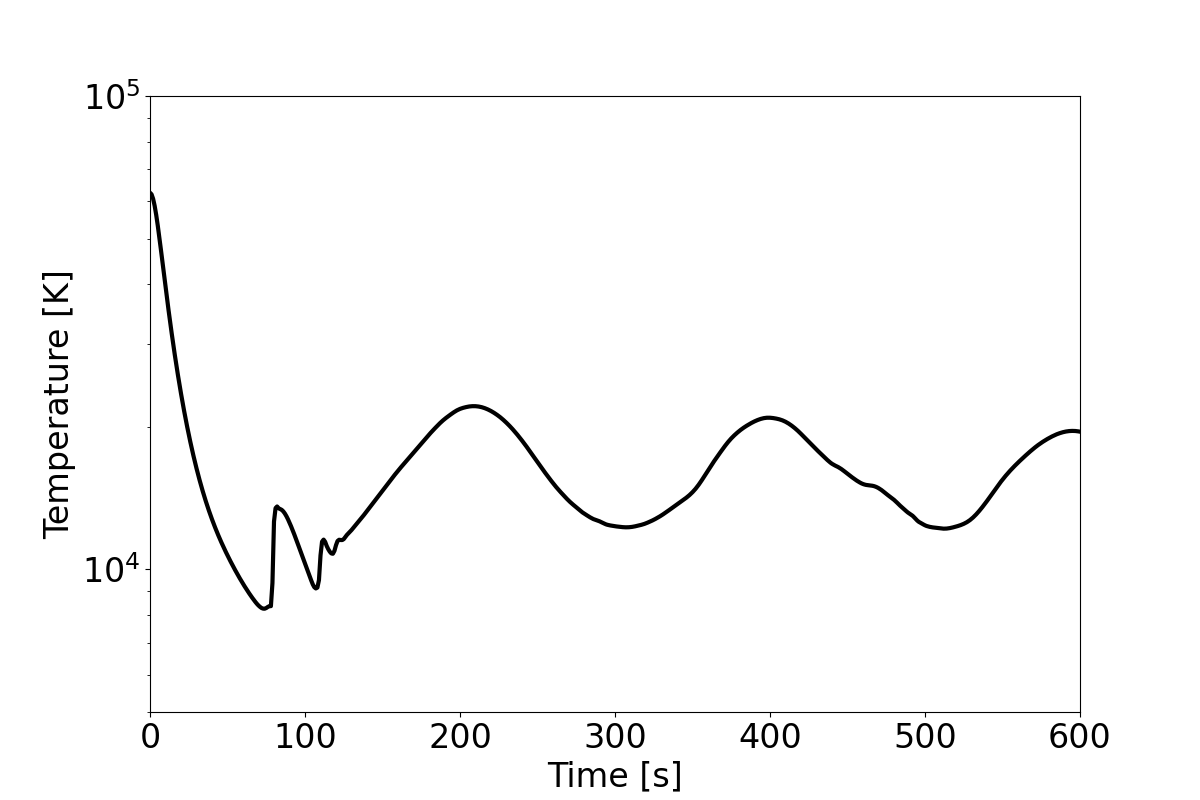}
     \includegraphics[width=8.0cm,height=5.5cm]{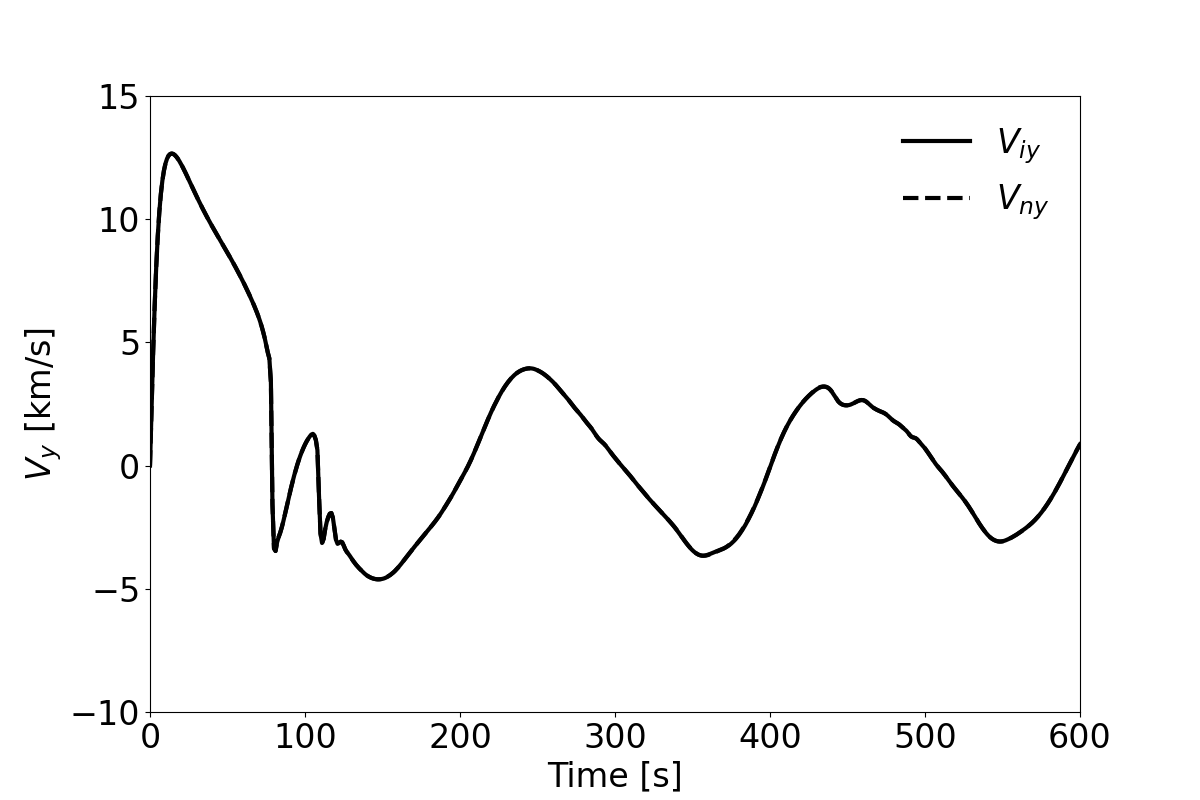}
      \caption{Time signatures of $\log T_{i}$ (left), and $V_{y_{i,n}}$  (right) collected at the point ($x=0$, $y=1.6$) Mm, corresponding to Run\#1.}
    \label{fig:Density_velocity_vs_time_Run1}
\end{figure*}

To identify the localized behavior of the jet at the null point, in Fig. \ref{fig:Density_velocity_vs_time_Run1}, we show the time series of $\log T_{i}$ (left), and vertical velocities $V_{y_{i,n}}$ (right) collected at the point $(x,y)=(0, 1.6)$ Mm, i.e., at the null point, corresponding to Run\#1. In these plots, we illustrate the behavior of the neutrals since we are analyzing their behavior at the point $(x,y)=(0, 1.6)$ Mm, located at the lower chromosphere, where the dynamic of neutrals is important. In $\log T_{i}$, we identify variations with a periodicity of about 200 s. On the right panel, we see that time signatures for vertical velocities of ions and neutrals behave alike and show an oscillatory behavior of about 200 s.

\begin{figure*}
      \includegraphics[width=8.0cm,height=5.5cm]{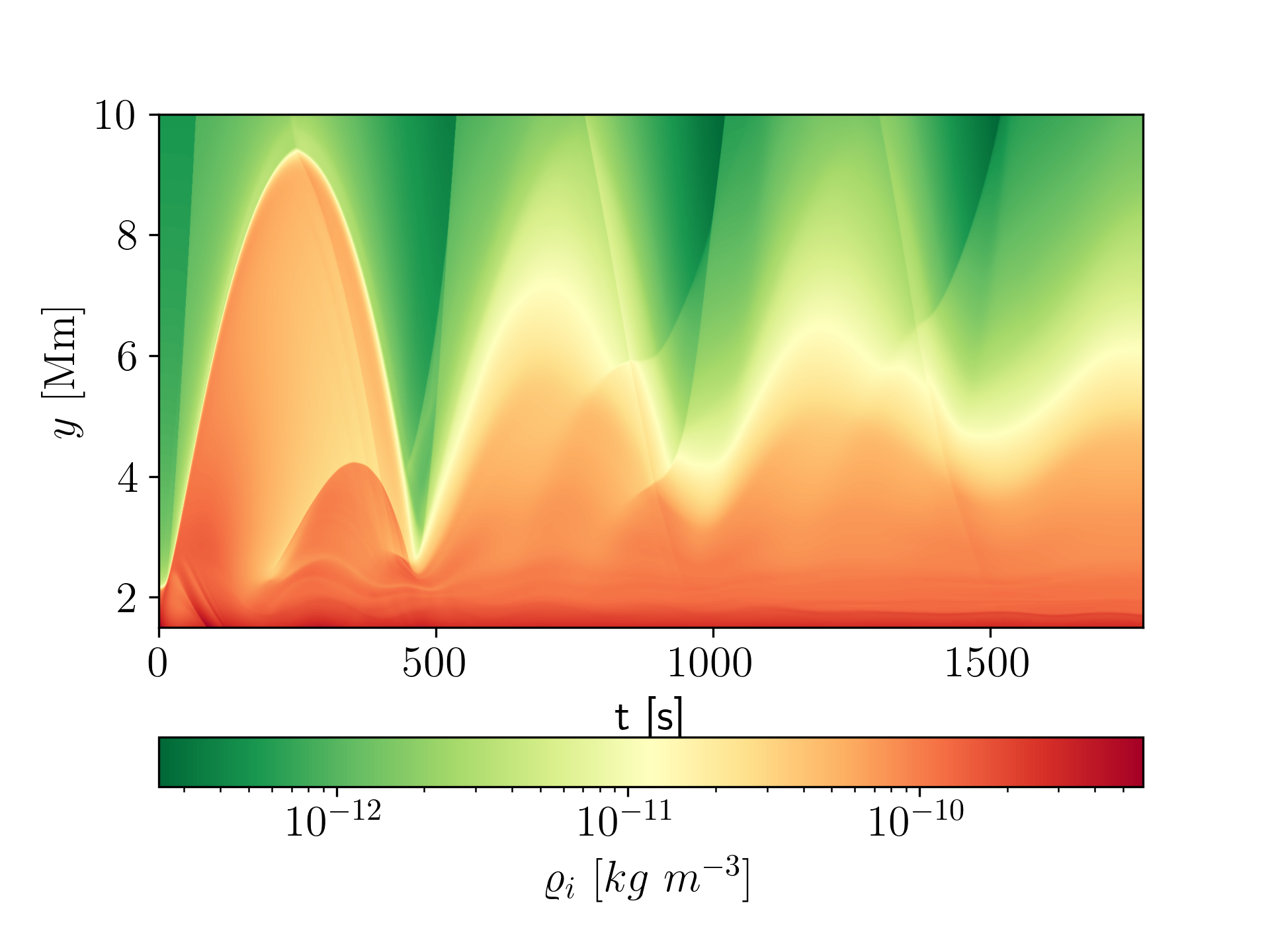}
 \includegraphics[width=8.0cm,height=5.5cm]{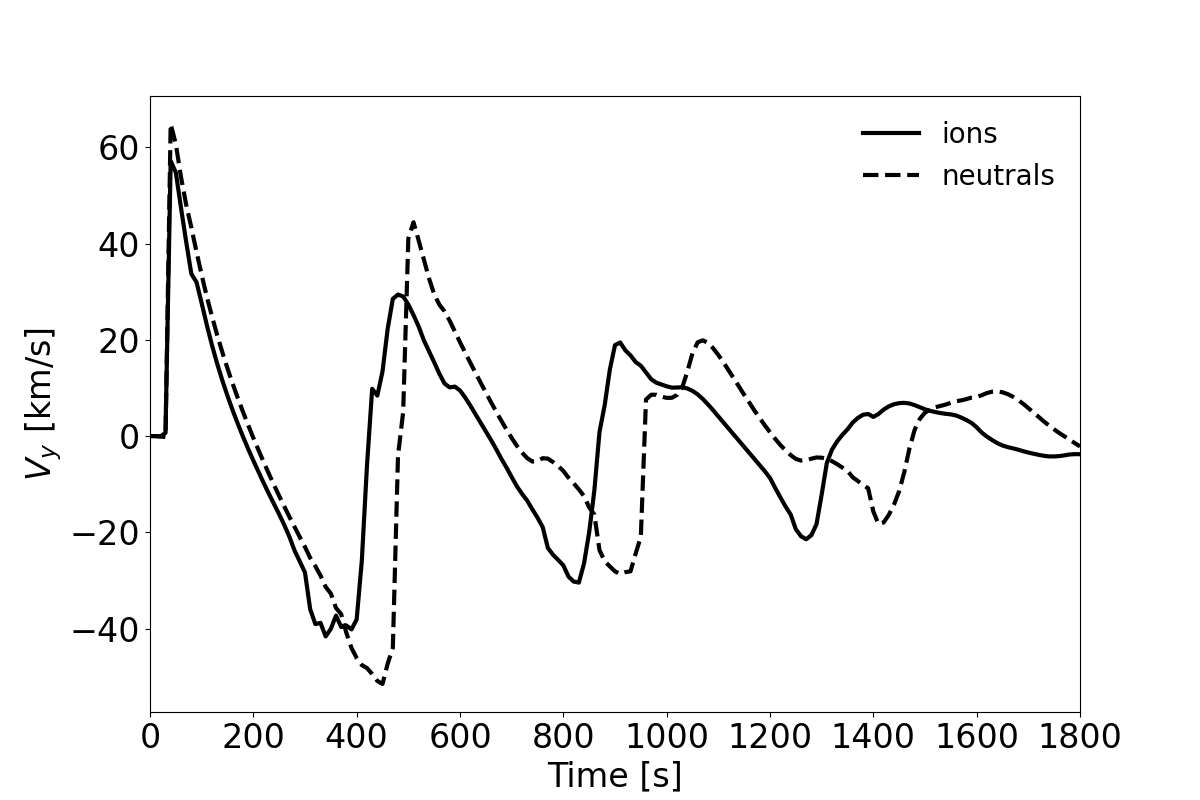}
     \includegraphics[width=8.0cm,height=5.5cm]{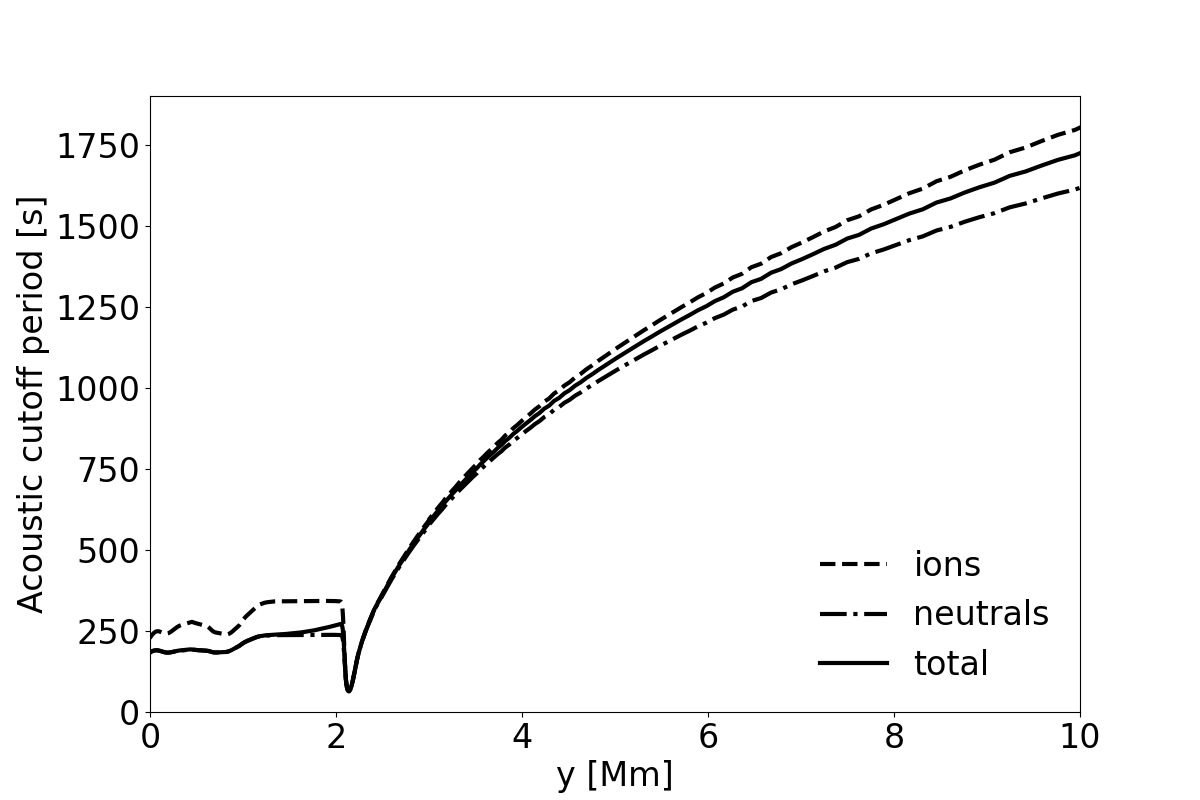}
      \caption{Top: Time-distance plots of $\log(\varrho_{i}(x=0,y))$ (left) and time series of $V_{y_{i,n}}$ collected at the point $(x=0,y=5)$ Mm (right). Bottom: The acoustic cut-off period, $P_{ac}$ for ion-neutral fluid (solid line), ionized fluid (dashed line), and neutrals (dot dashed line), as functions of height at initial time.}
    \label{fig:cutoff_period_sound_speed_vs_height_initial_time}
\end{figure*}

The plasma above the TR, for instance, at $y=5$ Mm for Run\#1, exhibits oscillations of about 450 s. This feature is initially schematized in the time-distance plot of the ionization fraction of Fig. \ref{fig:Distance-time_diagrams_If_vel_drift}, and confirmed in the time-distance diagram of the mass density of ions and the time series of the vertical velocity for ions and neutrals collected at $(x=0, y=5)$ Mm, as shown on the top-left and top-right of Fig. \ref{fig:cutoff_period_sound_speed_vs_height_initial_time} up to $t=1800$ s. These oscillations are due to secondary shocks through the process: excited pulse propagates upwards and steepens into shock due to the density stratification. This shock propagates from the chromosphere into the corona and lifts the chromospheric plasma behind. On the other hand, local chromospheric plasma starts to oscillate at local acoustic cut-off frequency after the propagation of pulses. In the linear limit, these oscillations are linear and decay as time progresses \citep{2009A&A...505..763K,2017A&A...601A..42P}. However, in the nonlinear limit, these oscillations lead to the excitation of quasi-periodic or secondary shocks. These shocks lift the chromospheric plasma with a periodicity, which depends on the initial amplitude of the pulse so that the stronger pulse leads to a longer periodicity \citep{2011A&A...529A..85Z}. 

We can compare the quasi-periodic oscillations of the secondary shocks shown on the top panels of Fig. \ref{fig:cutoff_period_sound_speed_vs_height_initial_time} to the acoustic cut-off period for a non-isothermal solar atmosphere in the linear regime, which following the work by \cite{Deubner&Gough_1984} is defined as follows: 
\begin{equation}
P_{ac} = \frac{4\pi\Lambda}{c_{s}\sqrt{1+2\frac{d\Lambda}{dy}}}.
\end{equation}
Here $c_{s}=\sqrt{\gamma p/\varrho}=\sqrt{\gamma \left(\varrho_{i}/\mu_{i}+ \varrho_{n}/\mu_{n}\right)\frac{k_{B}T}{m_{H}}/\varrho_{i}+\varrho_{n}}$ is the total sound speed, i.e., for ions plus neutrals, and $\Lambda=\left|p/\frac{dp}{dy}\right|$ is the single fluid scale height, with $p=p_{i}+p_{n}$ representing the total (ions plus neutrals) gas pressure. In the bottom panel of Fig. \ref{fig:cutoff_period_sound_speed_vs_height_initial_time}, we show the acoustic cut-off period $P_{ac}$ as function of height at the initial time of the simulations. The cut-off period is higher for ions than for neutrals, from $y=0$ to $y=2.1$ Mm, covering the photosphere to the chromosphere region. In particular, the values of $P_{ac}$ range from nearly 250 s at $y=0$ Mm, to around 350 s, from $y=1$ Mm to $y=2$ Mm. The cut-off period for neutrals is smaller remaining about 200 s up to the top chromosphere ($\sim 2$ Mm). The cut-off period for both ions and neutrals reaches its minimum value of about 50 s at the TR $\sim y=2.1$ Mm; higher up, the cut-off exponentially grows with height. In particular, for $y=y_{0}=1.6$ Mm, that is the initial position of pressure pulses for Run\#1, $P_{ac}$ attains a value of about $350$ s, as shown in the bottom of Fig. \ref{fig:cutoff_period_sound_speed_vs_height_initial_time}. The detected oscillation period of around 450 s  (see the top panels of Fig. \ref{fig:cutoff_period_sound_speed_vs_height_initial_time}) is longer due to the nonlinear consideration as shown by \citet{2011A&A...529A..85Z}. Our results are consistent with the classical models of rebound (secondary) shocks for spicules jet formation \citep[see, e.g.,][]{1982ApJ...257..345H,1988ApJ...327..950S,Muraswki&Zaqarashvilli_2010,Murawski_et_al_2011a}.

\subsection{Temperature and the inverted-Y shape jets}
\label{subsec:morphology_inversted-Y_shape jets}

\begin{figure*}
    \centering
         \includegraphics[width=5.0cm,height=5.0cm]{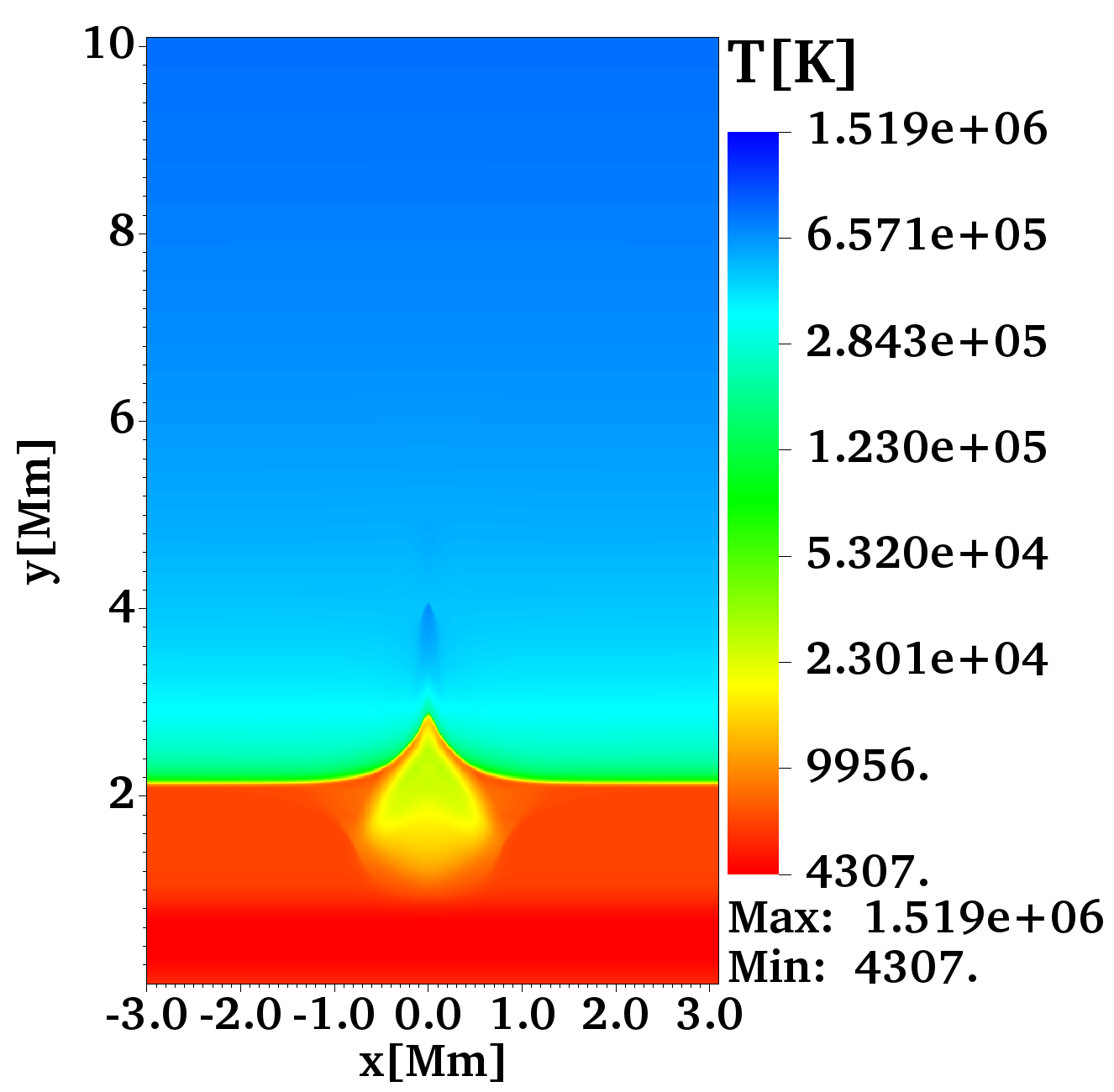}
     \includegraphics[width=5.0cm,height=5.0cm]{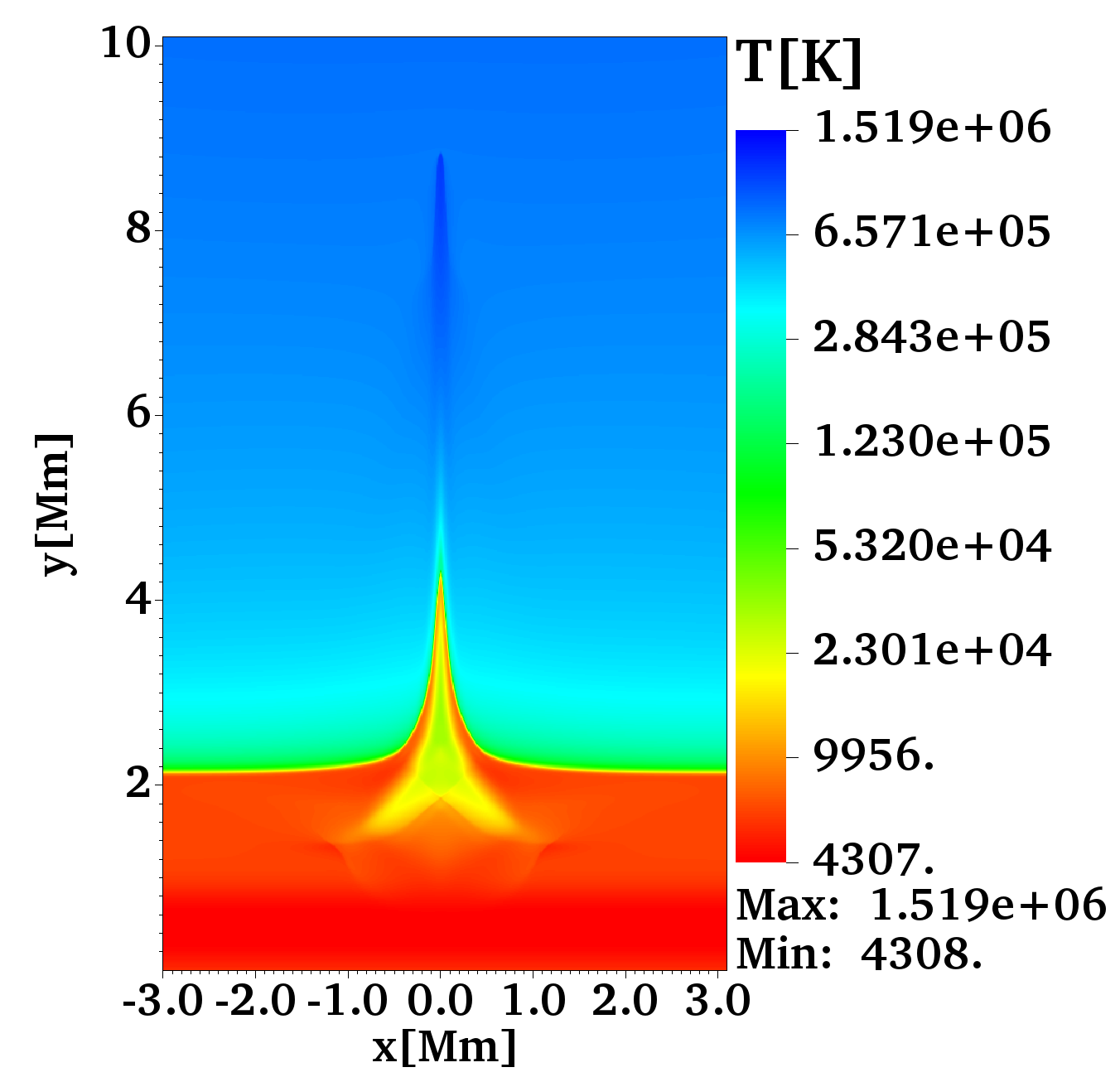}
      \includegraphics[width=5.0cm,height=5.0cm]{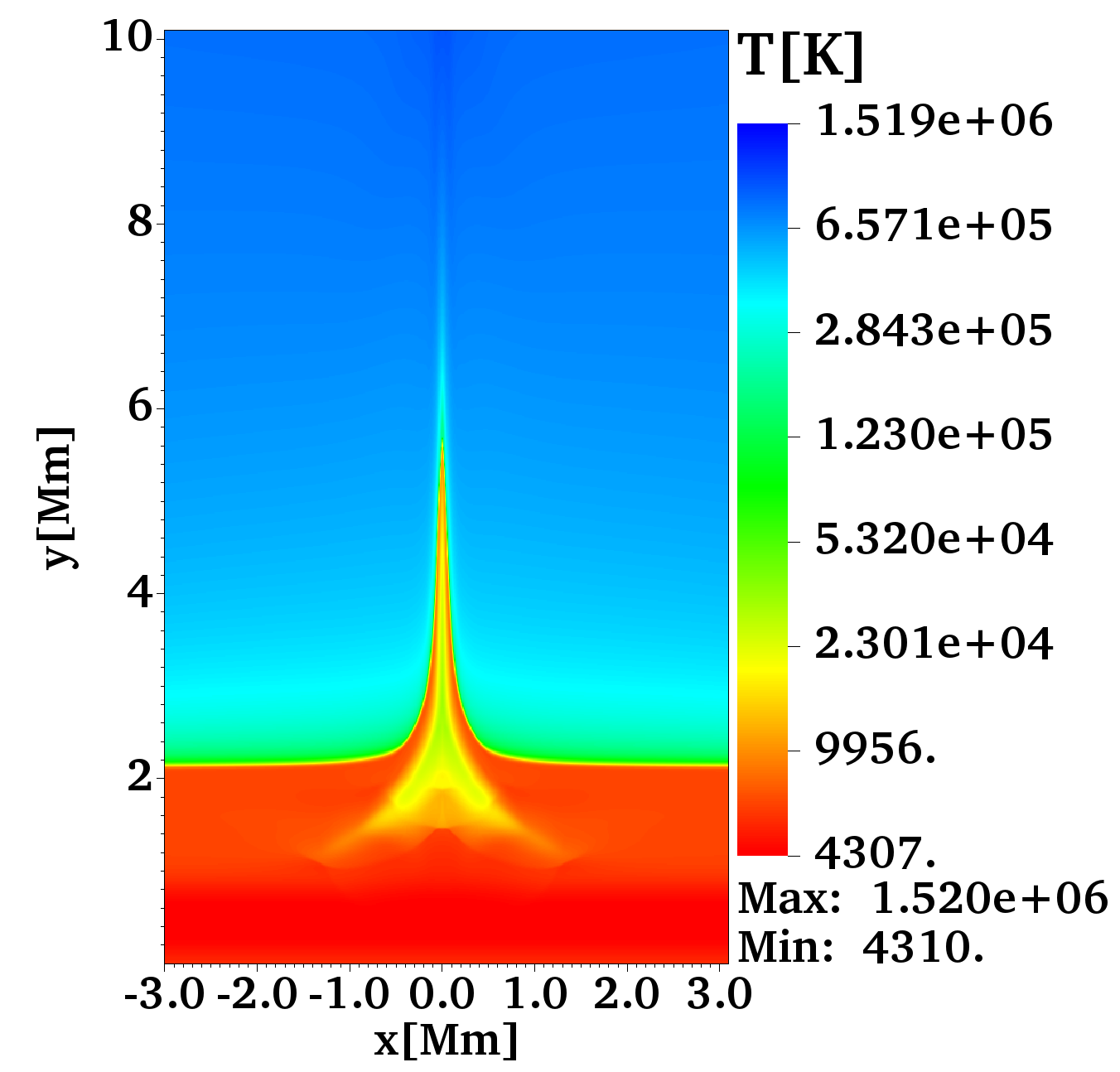}
    \caption{Ion temperature, $T_{i}$, maps in Kelvin, at time $t=30$ s (left), $t=60$ s (middle), and $t=90$ s (right), corresponding to Run\#1.}
    \label{fig:Temperature_maps_Run1}
\end{figure*}

To identify the evolution and behavior of ion temperature, which at $t=0$ s is equal to the neutral atom temperature, we show the temperature maps for three different instants of time $t=30, 60, 90$ s, corresponding to Run\#1 in Fig. \ref{fig:Temperature_maps_Run1}. On the left panel (corresponding to $t=30$ s), we see the two different parts of temperature: (i) the top part between the heights 3-4 Mm is made up of hot plasma of about $10^{5}$ K, which corresponds to the upward propagating first shock; (ii) the lower part below the height of 3 Mm shows colder ($\sim 10^{4}$ K) inverted-Y shape plasma, which corresponds to the chromospheric jet. In the middle panel (corresponding to $t=60$ s), the hot pulse propagates upwards and reaches the height of $y=9$ Mm, i.e., the mean velocity is around 170 km s$^{-1}$ (similar to local sound speed). At the same time, the cold jet reaches up to $y=4$ Mm, i.e., the mean velocity is 30 km s$^{-1}$ (similar to the speed of type I spicules). Finally, in the right panel ($t=90$ s), the hot pulse passed the considered heights, while the cold inverted-Y shape jet reached the height of $y=6$ Mm. It is also important to note that the plasma above the jet (green light color) is hotter ($\sim 10^{5}$ K) than inside (red and yellow colors). The temperature increase above the jet could be caused by the heating associated with ion-neutral collisions. This increase in temperature might be a novel result, which is connected to the two-fluid approximation. Indeed, Fig. \ref{fig:Distance-time_diagrams_If_vel_drift} (bottom-left) shows that $|V_{iy}-V_{in}|$ is large there, which results in deposition of the thermal energy.

From Fig. \ref{fig:Temperature_maps_Run1}, one can see that the jet shows inverted-Y shape after $t=60$ s. This inverted-Y shape is related to the magnetic null point as it is absent in similar simulations for a straight magnetic field.

The inverted-Y shape jets have been observed in various layers of the solar atmosphere and can be related to observational signatures of magnetic reconnection. For instance, \cite{Shibata_et_al_2007} found that Hinode observations show the ubiquitous presence of chromospheric anemone jets outside sunspots in active regions; these small jets had an inverted-Y shape, similar to the shape of x-ray anemone jets in the corona. Thus, they concluded that magnetic reconnection could occur at small spatial scales throughout the chromosphere. Additionally, \cite{Tian_et_al_2018} used the 1.6 m Goode Solar Telescope to detect prevalent reconnection through frequently occurring fine-scale jets in the H$\alpha$ line wings at light bridges of sunspots. They found that many jets had an inverted-Y shape, shown by models as typical of reconnection in a unipolar field environment.
Moreover, \cite{Nelson_et_al_2019} linked the inverted-Y shaped jets to numerous apparent signatures of magnetic reconnection in the solar photosphere. They used $H\alpha$ imaging data sampled by the Swedish Solar Telescope to investigate whether bidirectional flows can be detected within inverted-Y-shaped jets near the solar limb. They concluded that magnetic reconnection might cause bidirectional flows within inverted-Y-shaped jets and could be the driver of surges. Furthermore, in the context of solar jets and reconnection, \cite{2015Natur.523..437S} reported high-resolution X-ray and extreme-ultraviolet observations of jets that form in coronal holes at the Sun’s poles. They stated that contrary to the emerging-flux model, a miniature version of the filament eruptions that initiate coronal mass ejections drives the jet-producing reconnection.  

\subsection{Dynamics of magnetic field lines}
\label{sec:behavior_mag_field}

\begin{figure*}
    \centering
         \includegraphics[width=5.5cm,height=5.5cm]{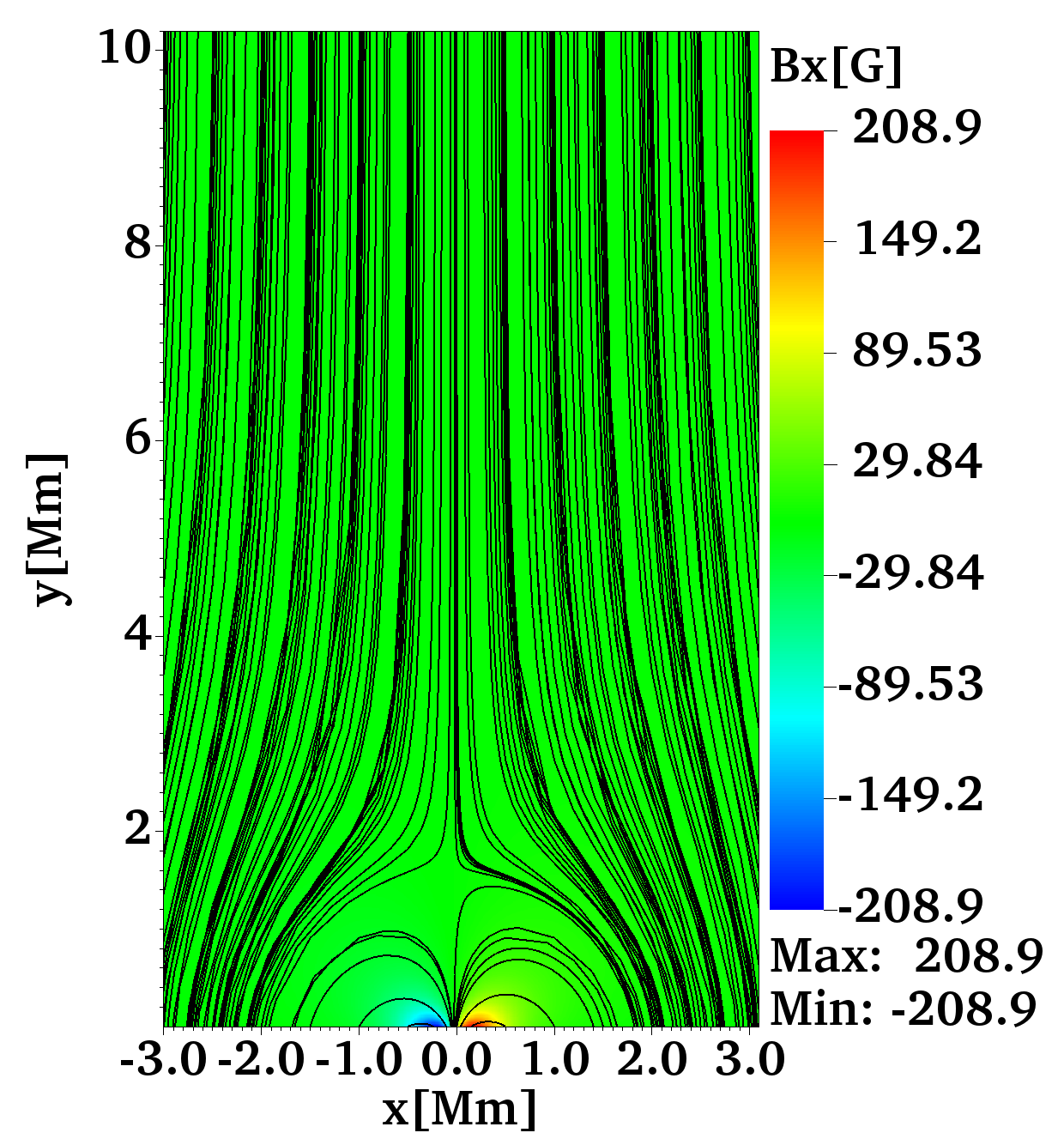}
      \includegraphics[width=5.5cm,height=5.5cm]{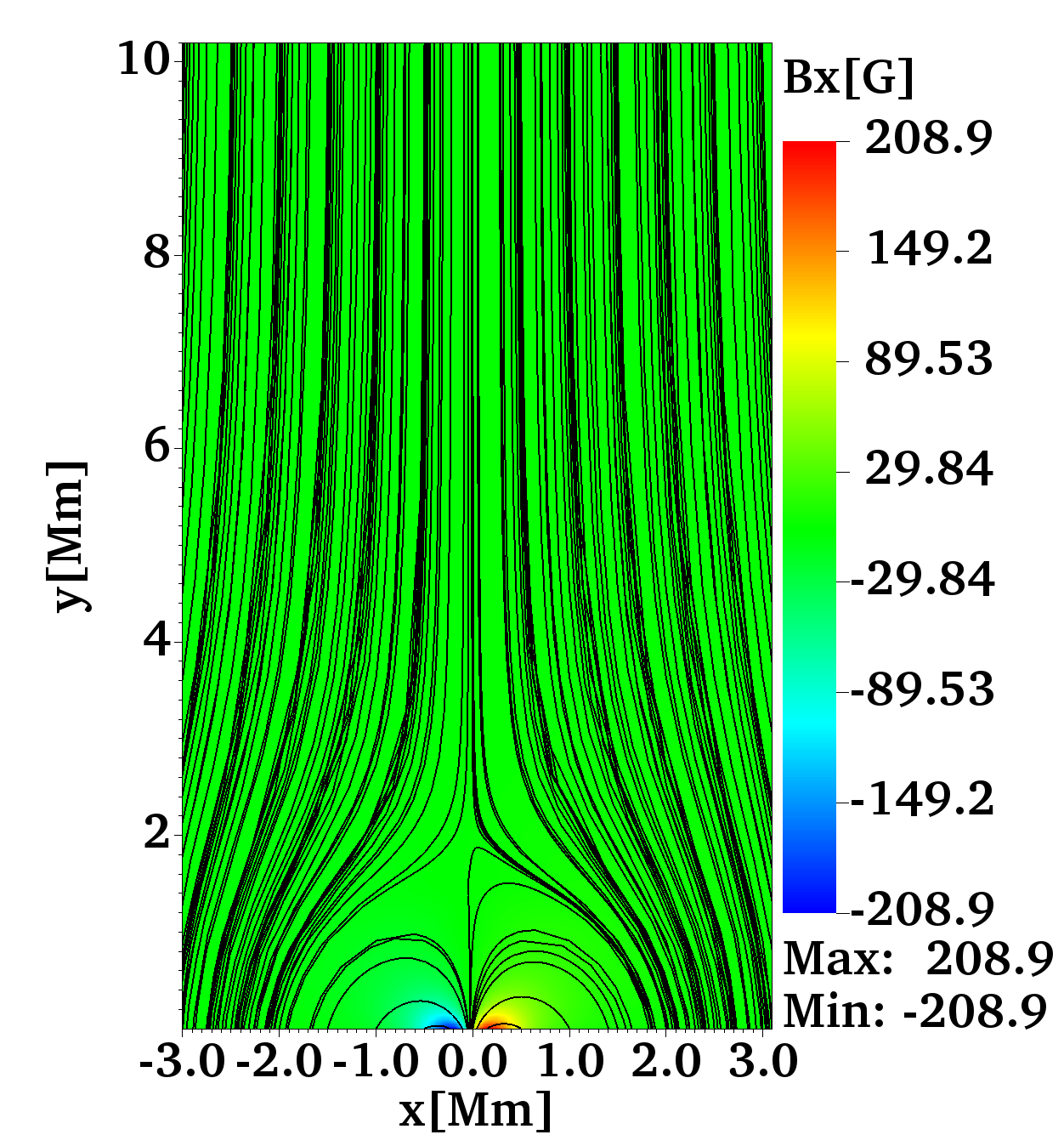}\\
       \includegraphics[width=5.5cm,height=5.5cm]{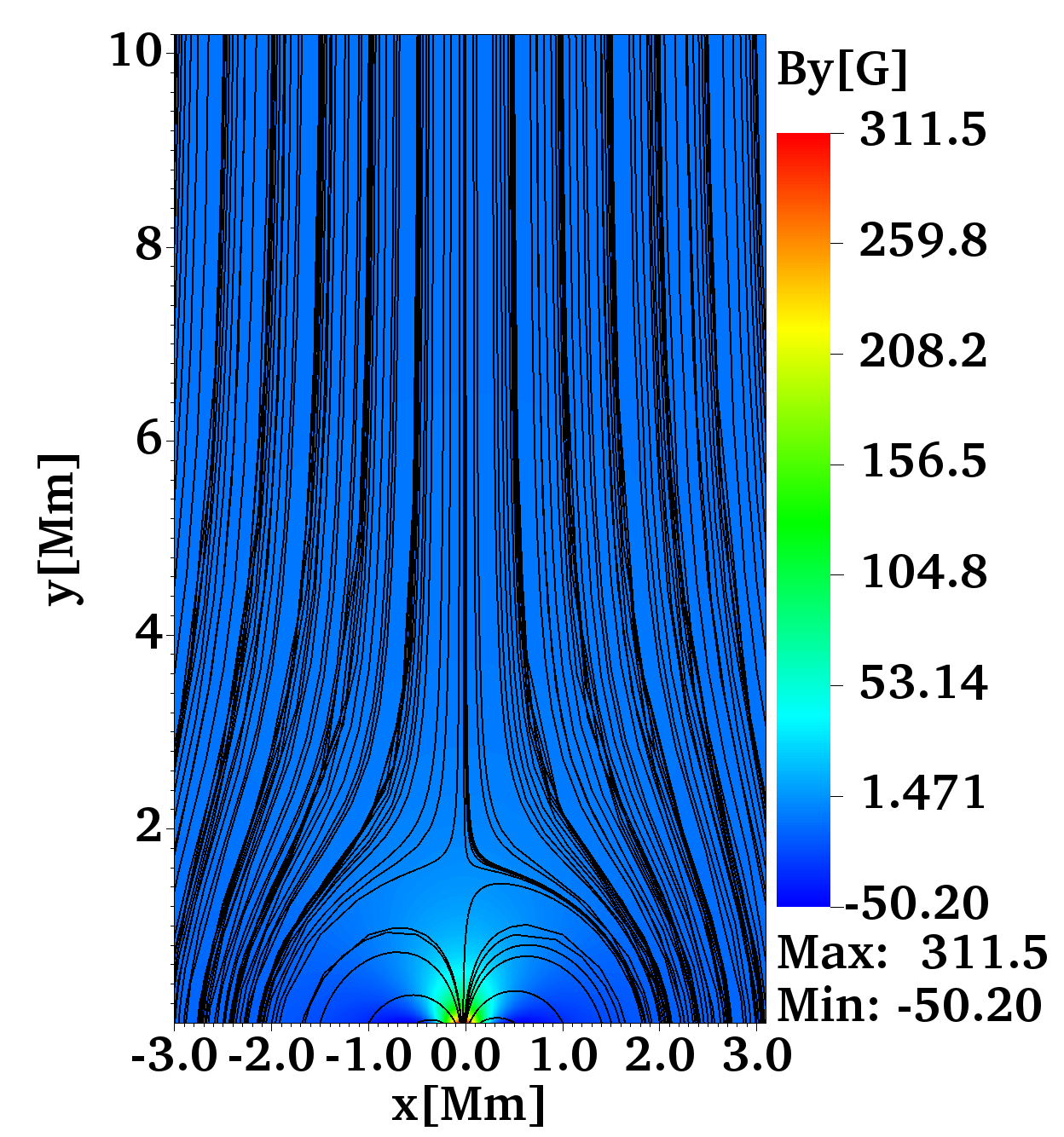}
      \includegraphics[width=5.5cm,height=5.5cm]{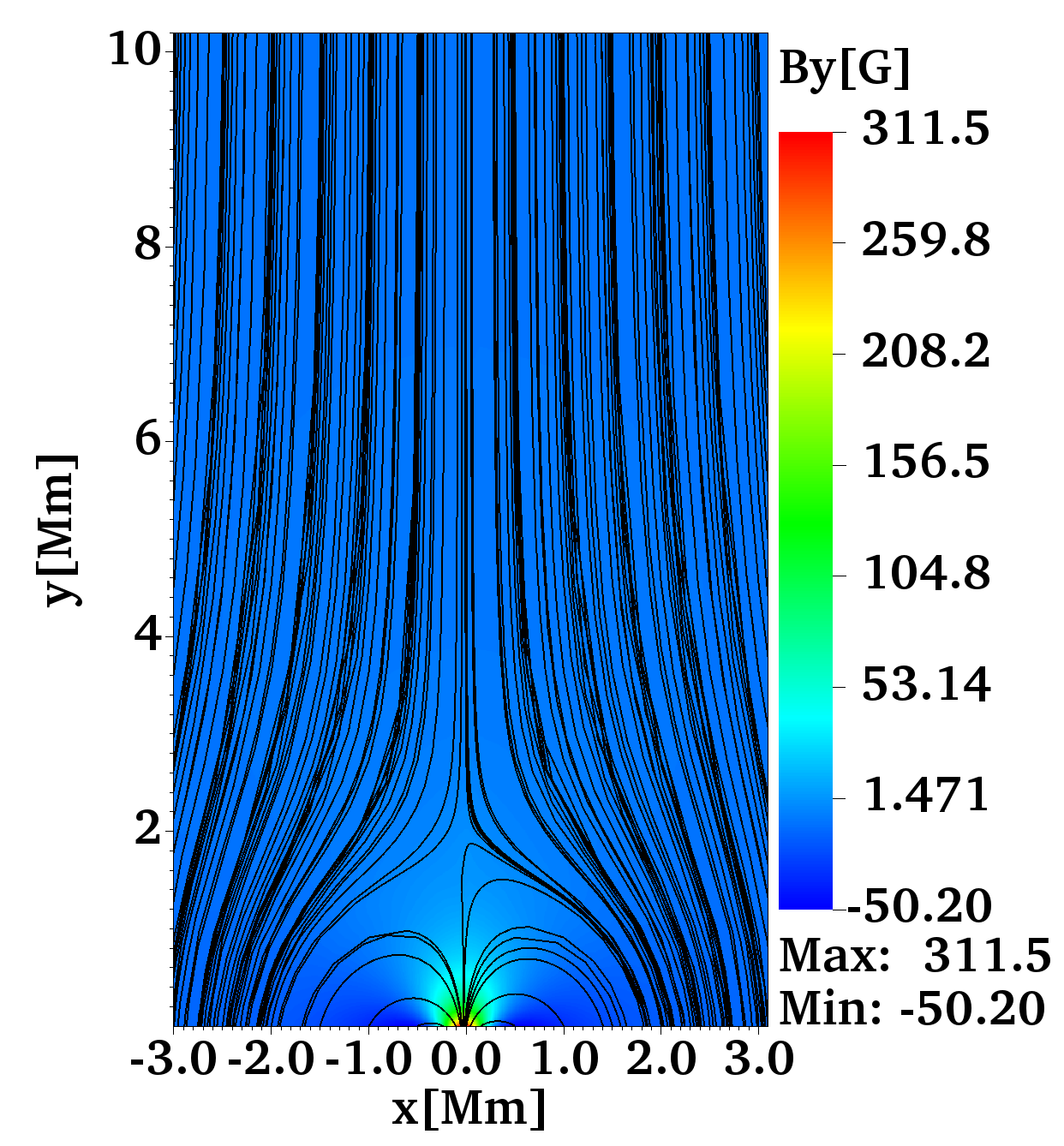}
    \caption{Maps of the horizontal, $B_{x}$ (top panels) and vertical, $B_{y}$, (bottom panels) components of the magnetic field, expressed in Gauss and overlaid by the magnetic field lines at times $t=0$ s (left panels) and $t=600$ s (right panels).}
    \label{fig:Magnetic_field_components_maps_field_lines_Run1}
\end{figure*}

In order to show the dynamics of the magnetic field during the simulations, on Fig. \ref{fig:Magnetic_field_components_maps_field_lines_Run1} we display initial and final maps of horizontal ($B_{x}$) and vertical ($B_{y}$) components corresponding to Run\#1. The upper panels of the figure illustrate $B_{x}$ at the initial ($t=0$ s) and final ($t=600$ s) times of simulation, which indicate that the configuration remains almost unchanged, with only a slight reconstruction of the field lines near the location of the null point ($y=1.6$ Mm). The lower panels show that the configuration of $B_{y}$ also did not change during simulation. This is because the generated pulse and jet propagate almost along the magnetic field lines and do not affect the lines themselves. On the other hand, the inverted-Y shape of the exciting jet is connected with the magnetic field configuration around the null point.

%
%
\section{Final comments and conclusions}
\label{sec:Conclusions}
We studied the formation and evolution of jets excited by the pulses in ion and neutral gas pressures. These pulses were launched from the null point of a potential magnetic arcade in a partially ionized and gravitationally stratified plasma. We used the two-fluid approximation in the numerical simulations, where electrons+ions and neutral atoms are recognized as two different fluids which are coupled by ion-neutral collisions. The two-fluid effects become significant near the time scales of ion-neutral collisions, much higher than $1$ Hz in the chromosphere. Nevertheless, for the processes with longer time scales in partially ionized plasma, two-fluid and single fluid (with Cowling diffusion) approximations can be applied without restriction. Both approximations give essentially the same results for the longer time-scales (see for example, \citep{Zaqarashvili_et_al_2011}). We chose the two-fluid approach because it is more appropriate for numerical simulations. 

For the magnetic field configuration, we used the 2D potential arcade model with a null point, which may result from the possible magnetic reconnection events in the solar atmosphere. First, we studied two different locations of the magnetic null points at the chromospheric ($Y= 1.6$  Mm) and the coronal ($Y = 2.2$ Mm) heights. Then, we launched  Gaussian pulses in ion and neutral gas pressures from the null points and followed their evolution in the stratified atmosphere. The evolution of the pulses and the whole system crucially depend on the initial location of the pulses (hence on the location of the null point). When the pulses were launched from $Y= 1.6$ Mm height, the following process was observed in numerical simulations: the pulses steepened into shocks, penetrating the corona and propagating with the mean velocity of coronal sound speed. The chromospheric cold and dense plasma followed the shock as a collimated jet and started to rise, reaching $8-9$ Mm height with the mean velocity of $20-30$ km s$^{-1}$. After a few hundreds of seconds, the dense and cold plasma of the jet started to flow down due to the gravitational attraction.
On the other hand, the pulses generated the quasi-periodic wake with secondary shocks. These shocks formed quasi-periodic upward motions of chromospheric partially ionized plasma in the form of collimated jets. However, for the pulses were launched at $Y= 2.2$ Mm height, no jets were observed as the excited waves did not steepen into shocks, and hence the chromospheric plasma did not rise into the corona.      
The important contribution of the magnetic structure is that the null points lead to the particular inverted-Y shape of the jets near the footpoints. Therefore, the observed inverted-Y shape of jets is probably connected to the reconnection events in the chromosphere rather than the corona. This point needs further study by observations and simulations.

Our simulations show that, according to our expectations, ions and neutrals evolve similarly from the high ion-neutral collision frequency. This fact means collisions couple the charged and neutral fluids on longer time scales than the collision time scales. On the other hand, the temperature around the jets significantly grows, which could be connected to the plasma heating due to the collisions. This temperature increase may explain the observed rapid disappearance of some jets from cold chromospheric spectral lines and their consequent appearance in hot transition region lines.

Another interesting consequence of the simulations is that the period of quasi-periodic upward motions of jets is longer than the acoustic cut-off period, which for the C7 model, ranges from about $250$ s at the photosphere ($y=0$ Mm) to $350$ s at the upper chromosphere ($y=2$ Mm). In particular, for near $y=y_{0}=1.6$ Mm, the cut-off period is approximately $350$ s, but the periodicity of plasma  oscillations over the TR is almost $450$ s, as shown on the left panel of Fig. \ref{fig:cutoff_period_sound_speed_vs_height_initial_time}. This is associated with the fact that the periodicity of quasi-periodic shocks depends on the initial amplitude of the pulse: stronger initial pulses lead to the longer periods of secondary shocks in the stratified atmosphere \citep{2011A&A...529A..85Z}.

In conclusion, the pressure pulses excited  (presumably by magnetic reconnection) near the magnetic null points in the chromosphere lead to the formation of quasi-periodic collimated jets, which rise into the hot solar corona. The lower part of these jets has an inverted-Y shape, which is a typical feature of some observed jets. In addition, the upper parts of these jets are heated, which could be connected to ion-neutral collisions. Future simulations, including variable ionization/recombination, may reveal more detailed features of the jet heating due to ion-neutral effects in partially ionized plasma.
\section*{Acknowledgements}
We thank the referee for the constructive comments that helped us to improve the manuscript. JJGA thanks to Investigadores por M\'exico-CONACYT (CONACYT Fellow), CONACYT LN 315829 (2021) and CONACYT-AEM 2017-01-292684 grants for partially support this work. The program "investigadores for M\'exico", project 1045 sponsor space Weather Service Mexico (SCIESMEX). We carry out the simulations in the facilities of ``Centro de Superc\'computo de Clima Espacial (CESCOM)"  part of the ``Laboratorio Nacional de Clima Espacial (LANCE)" at IGUM, Morelia, Michoacan, Mexico. The JOANNA code was developed by Darek W\'ojcik with some contribution from Piotr Wołoszkiewicz and Luis Kadowaki. KM's work was done within the framework of the National Science Centre (NCN) grant projects nos. 2017/25/B/ST9/00506 and 2020/37/B/ST9/00184. TVZ was supported by the Austrian Fonds zur F{\"o}rderung der Wissenschaftlichen Forschung (FWF) project P30695-N27 and and by Shota Rustaveli
National Science Foundation of Georgia (project FR-21-467). We visualize the simulation data using the VisIt software package \citep{Childs_et_al_2012}. 

\section*{Data Availability}
No new data were generated or analysed in support of this research.



\bibliographystyle{mnras}
\bibliography{Num_simulations_of_a_two_fluid_jet_at_magnetic_null_point_in_a_solar_arcade_MNRAS_2022} 



\bsp	
\label{lastpage}
\end{document}